\documentclass[12pt,a4paper]{article}
\usepackage[utf8]{inputenc}
\usepackage[lmargin=1in, rmargin=1in, tmargin=1in, bmargin=1in]{geometry}
\usepackage{setspace}
\usepackage{titling}
\usepackage{lipsum}
\usepackage{comment}
\usepackage{physics}
\usepackage{placeins}
\usepackage{enumerate}
\usepackage{longtable} %% to make tables span multiple pages
\usepackage{caption}
\usepackage{changepage}
\usepackage{appendix}

\allowdisplaybreaks
\RequirePackage{amsthm,amsmath,amsfonts,amssymb}
\usepackage{mathtools, nccmath}
\RequirePackage{natbib}
\RequirePackage[colorlinks,citecolor=blue,urlcolor=blue]{hyperref}
\RequirePackage{graphicx}

\usepackage{tikz}
\usetikzlibrary{shapes}
\usetikzlibrary{matrix}
\usepackage[normalsize]{subfigure}
\usetikzlibrary{backgrounds,automata}
\usepackage{float}

\usepackage{bbm}
\usepackage{mathrsfs}

\usepackage{makecell}
\def\X{{\boldsymbol{X} }}
\def\x{{\boldsymbol{x}}}
\def\G{{\boldsymbol G}}

\def\Z{{\boldsymbol Z}}

\def\one{\mathbbm{1}}

\newcommand{\Cov}{\operatorname{Cov}}
\newcommand{\Var}{\operatorname{Var}}
\newcommand{\E}{\operatorname{E}}
\newcommand{\defeq}{\vcentcolon=}
\newcommand{\pt}{\mathcal{P}_t}
\DeclareMathOperator{\Pro}{Pr}
\DeclareMathOperator{\logit}{logit}

\newcommand{\indep}{\rotatebox[origin=c]{90}{$\models$}}

\newcommand{\bigslash}{\mathrel{\smash{\raisebox{3ex}{\rotatebox{160}{\rule{0.5pt}{5ex}}}}}}

\theoremstyle{plain}
\newtheorem{prop}{Theorem}

\theoremstyle{remark}

\newtheorem{assu}{Assumption}

\begin{document}

\title{Counterfactual Slopes and Their Applications in Social Stratification}
\author{Ang Yu\thanks{Department of Sociology, University of Wisconsin-Madison. Email: ayu33@wisc.edu}, Jiwei Zhao\thanks{Department of Statistics and Department of Biostatistics \& Medical Informatics, University of Wisconsin-Madison}}
\date{September 21, 2024}

\setstretch{1.5}

\maketitle
\begin{abstract}
This paper addresses two prominent theses in social stratification research, the great equalizer thesis and Mare's (1980) school transition thesis. Both theses describe the role of an intermediate educational transition in the association between socioeconomic status and an outcome variable. However, the descriptive regularities of the two theses may be driven by differential selection into the intermediate transition, which prevents the two theses from having substantive interpretations. We propose a set of novel counterfactual slope estimands, which capture these theses under hypothetical interventions that eliminate the differential selection. We thereby construct selection-free tests for these theses. Compared with the existing literature, we are the first to explicitly provide nonparametric causal estimands, which enable us to conduct more principled analysis. We are also the first to develop flexible, efficient, and robust estimators for the two theses based on efficient influence functions. We apply our framework to a nationally representative dataset in the United States and re-evaluate the two theses. Findings from our selection-free tests suggest that the descriptive regularities are misleading for the substantive interpretation of the great equalizer thesis, but not for that of the school transition thesis. Additionally, the counterfactual slopes also provide a new framework for evaluating the inequality impacts of policy interventions. 
\end{abstract}

\section{Introduction}
\subsection{Two theses in social stratification}
In this paper, we evaluate two prominent theses in the research on social stratification and mobility, the great equalizer (GE) thesis and the school transition (ST) thesis. Both theses describe the role of an intermediate educational transition in the association between family social-economic status (SES) and an outcome variable.
The GE thesis describes the empirical regularity that the association between SES background and one's own SES attainment (outcome variable) is weaker among college graduates than non-college graduates \citep{hout_status_1984, hout_more_1988}.
In this case, college completion is the intermediate transition.
The ST thesis is the empirical regularity that the SES disparity in students’ likelihood of transitioning to a higher level of education decreases as they advance through the educational system \citep{mare_social_1980}.
Here, the outcome is whether the student made a higher-level educational transition, and the intermediate transition is a lower-level transition that is a prerequisite for making the higher-level transition.
Both theses are documented for multiple birth cohorts and multiple countries \citep{breen_change_2004, breen_social_2007, breen_explaining_2007, torche_is_2011, pong_trends_1991, shavit_persistent_1993}.

These descriptive regularities have been given substantive explanations. For the GE thesis, the difference in the association between SES origin and destination by college graduation may be explained by different labor market conditions facing college graduates and non-graduates. Labor markets for college graduates may be more meritocratic, where employers may be less influenced by employees' social origins \citep{breen_explaining_2007}. For the ST thesis, it is possible to explain the declining influence of family background over educational transitions from a life course perspective.
With increasing age, students may be increasingly able to make educational decisions on their own and less influenced by their parents' SES \citep{muller_social_1993, stolzenberg_educational_1994}. These substantive explanations are inherently causal, as they imply that had different individuals were counterfactually intervened to make the intermediate transition, the descriptive regularities would persist, because the substantive mechanisms should play out regardless of who make the transition (see, e.g., \citet{shavit_persistent_1993}, p.9; \citet{hout_more_1988}, p.1384).

\subsection{Complications in interpretation}
The substantive interpretations of the two theses are, however, complicated by the possibility of differential selection into the intermediate educational transitions. Sociologists have surmised that educational transitions are more of a rational decision based on expected outcomes and less of a cultural norm among the disadvantaged than among the advantaged individuals \citep{mare_social_1980, hout_social_2012}. Consequently, for young people of lower origins, only those who are particularly able and motivated might make the transitions. Meanwhile, the transitions may be relatively more uniformly attained by young people of higher origins. In fact, even back in 1965, \citet[][p.43-7]{bourdieu_rapport_1965} have noted that, compared with middle-class children, working-class children have undergone a greater degree of selection when they reach a higher level of education.

In the context of the GE thesis, if among college graduates, socioeconomically disadvantaged students are more highly selected on characteristics such as ability and motivation than their advantaged peers, then this differential selection may partly counteract the influence of family background among college graduates, creating a weaker association between family background and SES attainment. Similarly, for the ST thesis, socioeconomically disadvantaged students who have made the intermediate transition may be more selected, too, again leading to an attenuated association between family background and making the higher-level transition.

The implication of differential selection into the intermediate transition for the association between family background and the outcome can be understood from the perspective of collider bias \citep{hernan_structural_2004,elwert_endogenous_2014}. Family background and pre-transition covariates such as ability and motivation both contribute to the probability of achieving the transition, and the pre-transition covariates also affect the outcome. Consequently, by conditioning on achieving the transition, which is a collider, the association between family background and the pre-transition covariates is distorted. In this case, the association is likely to be spuriously weakened.\footnote{Under a linear structural equation model, this is necessarily true if SES background and a pre-transition covariate  positively and additively affect the transition \citep{pearl_linear_2013}.} Finally, the weakened association between family background and the pre-transition covariates results in a weakened association between family background and the outcome, creating the empirical regularities of the GE and ST theses \citep[see the directed acyclic graphs in][]{zhou_equalization_2019}.

If these differential selection patterns hold, then the substantive explanations outlined above may not be supported as much as the descriptive regularities appear to suggest. In particular, if characteristics of the individuals who make the intermediate transition explains the descriptive regularities, counterfactually assigning different individuals to receive the intermediate transition would alter the descriptive regularities, which contradicts the substantive explanations. 
Whether differential selection drives the descriptive regularities also has policy implications.
For the GE thesis, if a college degree does not causally reduce the intergenerational association in SES attainment, then intervening to expand college education might not achieve the level social mobility currently observed among college graduates. Similarly, for the ST thesis, expanding access at a lower level of education might not lead to as much overall educational equity as might be expected based on the descriptive regularity.

\subsection{Our approach}

In this paper, we propose a novel statistical approach designed to test the two theses in a manner free from selection bias. This allows us to eliminate the spurious impact of differential selection and assess the substantive interpretations of the two theses.
To begin with, we introduce the following notation. Let $G$ be SES background, $Y$ be the outcome variable, and $D$ be the intermediate transition, which is binary.
Note that $Y$ could be either a continuous variable (SES in adulthood in the GE thesis) or a binary variable (an educational transition at a level higher than $D$ in the ST thesis). Finally, $Y_d$ denotes the potential outcome of $Y$ under the external assignment of a transition status $d$ (0 or 1) \citep{rubin_estimating_1974}. Although all variables are individual-level quantities, we omit the individual subscript $i$ in most cases to unburden notation. 

The descriptive regularities summarize conditional expectations in the form of  $\E(Y \mid D=d, G)$, the observed mean of $Y$ given $G$ and a transition status $d$. In contrast, to test the two theses in a selection-free manner, we propose an approach that focuses on $\E(Y_d \mid G)$, the counterfactual mean of $Y$ given $G$ under the external assignment of a transition status $d$. Under external assignment of the transition, individuals no longer self-select into the transition. Hence, by comparing $\E(Y_d \mid G)$ across $G$ values, we will be able to characterize the relationship between $Y$ and $G$ under $D=d$ when differential selection is taken out of the picture. 

As $G$ is a continuous measure of income in this study, we need to define a summary measure of the relationship between $Y$ (or $Y_d$) and $G$ to make conceptual summary and statistical testing possible. We do so by borrowing the construct of slope in regression models. Depending on whether the outcome is continuous or binary, we consider slopes in both linear and logit forms. Combining the potential outcome framework with the slope formulation, we propose a set of counterfactual slope estimands which provides unified selection-free tests for both the GE and ST theses.

The rest of this paper is organized as follows. Section 2 formally introduces the counterfactual slope estimands, defines our selection-free tests, and lays out the assumptions we use to identify the estimands. Section 3 connects our approach to prior work in sociology on the GE and ST theses and discusses what make the latter inadequate. In Section 4, we develop doubly robust and semiparametrically efficient estimators for our estimands. Section 5 presents the empirical study on the GE and ST theses we conduct using our new framework. Section 6 discusses two alternative formulations of the tests for the ST thesis. Finally, Section 7 concludes with possible extensions. R Code for the empirical analysis is available at \url{https://github.com/ang-yu/Counterfactual_slope}.

\section{Counterfactual slopes and selection-free tests}\label{sec:proposal}

We first define two different counterfactual slope estimands corresponding to different types of outcome variables in the GE and ST theses.
For the GE thesis, where $Y$ is continuous, we define a linear counterfactual slope estimand for each transition status $d$:
\begin{equation*}
    \xi_{\text{linear}}(d) \vcentcolon= \frac{\Cov(Y_d, G)}{\Var(G)}=\frac{\Cov[\E(Y_d \mid G), G]}{\Var(G)},
\end{equation*}
which is the slope coefficient in the linear projection of $Y_d$ or $\E(Y_d \mid G)$ onto the space spanned by
$G$, namely the slope coefficient in the population linear regression of $Y_d$ on G. When $G$ is binary with the support $\{0,1\}$, $\xi_{\text{linear}}(d)$ becomes $\E(Y_d \mid G=1)-\E(Y_d \mid G=0)$, which is considered by \citet{naimi_mediation_2016} and \citet{lundberg_gap-closing_2024}. For the ST thesis where $Y$ is binary, we define a logit-transformed counterfactual slope estimand for each $d$:
\begin{equation*}
    \xi_{\text{logit}}(d) \vcentcolon= \frac{\Cov \left\{ \logit[\E(Y_d \mid G)], G \right\}}{\Var(G)},
\end{equation*}
where $\logit(x) \defeq \log\frac{x}{1-x}$. This is the slope coefficient in the linear projection of $\logit[\E(Y_d \mid G)]$ on the space spanned by $G$. If $\logit[\E(Y_d \mid G)]$ is linear in $G$, i.e., $\logit[\E(Y_d \mid G)]$ takes the form $\alpha + \beta G$,\footnote{Or equivalently, $\E(Y_d \mid G)=\frac{\exp(\alpha+\beta G)}{1+ \exp(\alpha+\beta G)}$. This logistic model of the potential outcome is also considered in \citet{karlson_marginal_2023}, although for different estimands.} then $\xi_{\text{logit}}(d)$ will have two additional interpretations. First, it will be the slope coefficient ($=\beta$) in the population logistic regression of $Y_d$ on $G$. Second, it will also be the change in $Y_d$ associated with a unit increase in $G$ on the log odds ratio scale, i.e., 
$$\log \left[ \frac{\E(D \mid G=g+1)}{1-\E(D \mid G=g+1)} 
\bigslash \frac{\E(D \mid G=g)}{1-\E(D \mid G=g)} \right].$$
Importantly, however, the slope estimand $\xi_{\text{logit}}(d)$ is defined nonparametrically and does \emph{not} impose the linearity of $\logit[\E(Y_d \mid G)]$. It is only the two additional interpretations that rely on the linearity.

These counterfactual slopes provide succinct summaries of the counterfactual relationship between $Y$ and $G$ under the universal assignment of transition status $d$. Under the hypothetical intervention of assigning $d$, all (potentially differential) selection into the transition is eliminated. Selection-free tests of the GE and ST theses can thereby be constructed. Note that the potential outcomes are defined with respect to only $D$, not $G$. This is because we are not concerned with the causal effect of family background. Instead, we are interested in the descriptive disparity in an outcome based on family background. 

The slope estimands do not impose any assumption on the functional form of the counterfactual relationship between $Y$ and $G$. We use the slopes to summarize the potentially complicated function $\E(Y_d \mid G)$. This is not only necessary for constructing statistical tests but also consistent with the conceptual narratives of the two theses—the GE and ST theses are about the overall strength of the (conditional or counterfactual) association between $Y$ and $G$, not the granular functional form of their relationship.

We have defined the counterfactual slopes using deterministic interventions, i.e., universal assignments of a specific transition status. The counterfactual slopes can also be interpreted in terms of a stochastic intervention, which more directly maps onto the goal of hypothetically eliminating differential selection into the transition. Under the stochastic intervention notion \citep{didelez_direct_2006, geneletti_identifying_2007}, $R(D \mid  G=g)$ is defined to be a randomly drawn value of $D$ from those with $G=g$. By this definition, we have
\begin{equation*}
    \E(Y_d \mid G=g) = \E \left[Y_{R(D \mid G=g)} \mid G=g, R(D\mid G=g)=d \right].
\end{equation*}
Hence, the $d$-specific counterfactual slopes represent the counterfactual association between $Y$ and $G$ among people who happen to be assigned $d$ in the hypothetical intervention to randomly re-assign transition status within $G$ levels.
Intuitively, this is because the mean counterfactual outcome of a randomly selected group of individuals is naturally equal to the mean counterfactual outcome of all individuals.
Unlike the deterministic intervention, this intervention does not change the transition rate at any level of $G$ but only eliminates selection into the transition within each level of $G$.

\subsection{Selection-free tests}
We define our selection-free tests of the two theses using the counterfactual slopes. The selection-free test statistic (with a slight abuse of terminology, we refer to estimands, not estimates, as test statistics) for the GE thesis is
\begin{equation}
    \xi_{\text{GE, selection-free}} \defeq \frac{\Cov(Y_0, G)}{\Var(G)}-\frac{\Cov(Y_1, G)}{\Var(G)} = \frac{\Cov[\E(Y_0 \mid G), G]}{\Var(G)}-\frac{\Cov[\E(Y_1 \mid G), G]}{\Var(G)}. \label{equ:free_GE}
\end{equation}
This test statistic contrasts the association between SES attainment and SES background under two hypothetical scenarios where college completion or non-completion is uniformally assigned. Importantly, there is no selection into college completion under both hypothetical scenarios. The null hypothesis of the selection-free test for the GE thesis is then $\xi_{\text{GE, selection-free}}=0$. 
%The alternative hypothesis is $\xi_{\text{GE, selection-free}}>0$, i.e., the association between SES background and attainment is weaker under the assignment of college completion than under the assignment of non-completion.
If $\xi_{\text{GE, selection-free}}$ is estimated to be greater than 0, and the null hypothesis is rejected, then the association between SES background and attainment is weaker under the assignment of college completion than under the assignment of non-completion, and we have evidence that is free of differential selection to support the substantive interpretation of the GE thesis.\footnote{The substantive interpretation of the GE thesis is supported only when $\xi_{\text{GE, selection-free}}>0$, not when $\xi_{\text{GE, selection-free}}<0$. Nonetheless, we opt to conduct a two-sided test, which is more conservative than a one-sided test, given that it is theoretically implausible to have $\xi_{\text{GE, selection-free}}<0$. We proceed similarly for other tests below.}

The selection-free test statistic for the ST thesis is
\begin{equation}
    \xi_{\text{ST, selection-free}} \defeq \frac{\Cov \left\{ \logit \left[\E(D \mid G) \right], G \right\}}{\Var(G)} - \frac{\Cov \left\{ \logit \left[\E(Y_1 \mid G) \right], G \right\}}{\Var(G)}. \label{equ:free_ST}
\end{equation}
This test statistic contrasts the factual relationship between $D$ (the lower-level transition) and $G$ (SES background) with the counterfactual relationship between $Y_1$ and $G$, i.e., the relationship between $Y$ (the higher-level transition) and $G$ when selection into the lower-level transition is purged. Note that the test statistic here is structurally different from the one for the GE thesis. For the ST thesis, the intermediate transition $D$ is not only used to define the potential outcome $Y_1$, but its factual relationship with $G$ is also part of the test statistic.
%And the alternative hypothesis is $\xi_{\text{ST, selection-free}}>0$, i.e., the association with SES background is weaker at the higher-level transition even if everyone had made the lower-level transition. 
Here, the null hypothesis is $\xi_{\text{ST, selection-free}}=0$. If $\xi_{\text{ST, selection-free}}$ is estimated to be greater than 0, and the null hypothesis is rejected, then the association between educational transition and SES background would be weaker at the higher level even if everyone had made the lower-level transition, which is evidence to support the substantive interpretation of the ST thesis.

\subsection{Identification}
\label{sec: assu}
We let $\X$ be a vector of baseline covariates. For ease of notation, we assume $ G \in \X $.
Our identification strategy relies on three standard assumptions for causal identification in observational studies:
\begin{assu}[Conditional ignorability]
     $Y_d \indep D \mid \X=\x, \forall d, \x$. \label{assu1}
\end{assu}

\begin{assu}[Stable Unit Treatment Values Assumption (SUTVA)]
    $Y=D Y_1 + (1-D) Y_0$.\label{assu2}
\end{assu}

\begin{assu}[Overlap]
    $0 < \E(D \mid \x) <1, \forall \x$.  \label{assu3}
\end{assu}
The conditional ignorability assumption requires the potential outcomes to be independent of transition status $D$ given observed confounders. The SUTVA \citep{rubin_randomization_1980} states that the same transition assignment is always associated with the same potential outcome for any individual, and the potential outcome of any individual is not affected by another individual' transition assignment. And the overlap assumption means that the probability of achieving the transition is bounded away from 0 and 1 conditional on the confounders.
Under these assumptions, we can identify the linear counterfactual slope $\xi_{\text{linear}}(d)$ as
\begin{equation*}
    \xi_{\text{linear}}(d) = \frac{\E[G \mu(d,\X)]-\E[\mu(d,\X)]\E(G)}{\Var(G)} ,
\end{equation*}
where $\mu_d(\X) \vcentcolon= \E(Y \mid D=d, \X)$. Similarly, the logit counterfactual slope is identified as
\begin{equation*}
    \xi_{\text{logit}}(d) = \frac{\E \left\{ G \cdot \logit\{\E[\mu(d,\X) \mid G] \} \right\} - \E\left\{ \logit\{\E[\mu(d,\X) \mid G] \} \right\}\E(G)}{\Var(G)}.
\end{equation*}

\section{Connection with the literature}

\subsection{Descriptive tests}
We also formalize two descriptive tests that capture the original empirical regularities of the two theses.
For the GE thesis, we define the descriptive test statistic as the difference between two linear factual slopes:
\begin{equation}
    \xi_{\text{GE, descriptive}} \defeq \frac{\Cov[\E(Y \mid D=0, G), G]}{\Var(G)}-\frac{\Cov[\E(Y \mid D=1, G), G]}{\Var(G)}. \label{equ:desc_GE}
\end{equation}
This difference does not contain potential outcomes and contrasts factual associations between SES background and SES attainment for college graduates and non-graduates. Analogous to the selection-free test, the null hypothesis is $\xi_{\text{GE, descriptive}}=0$. If $\xi_{\text{GE, descriptive}}$ is estimated to be greater than 0, and the null hypothesis is rejected, we will have descriptive evidence to support the GE thesis, as the intergenerational SES association is weaker among college graduates.
Similarly, for the ST thesis, the descriptive test statistic is the difference between two logit factual slopes:
\begin{equation}
    \xi_{\text{ST, descriptive}} \defeq \frac{\Cov \left\{ \logit[\E(D \mid G)], G \right\}}{\Var(G)} - \frac{\Cov \left\{ \logit[\E(Y \mid D=1, G)], G \right\}}{\Var(G)}. \label{equ:desc_ST}
\end{equation}
Again, the null hypothesis is $\xi_{\text{ST, descriptive}}=0$. If $\xi_{\text{ST, descriptive}}$ is estimated to be greater than 0, and the null hypothesis is rejected, we will have descriptive evidence for the ST thesis, as the association of SES background with making a lower-level school transition is stronger than that with making a higher-level transition conditional on having made the lower-level transition.

Naturally, the descriptive test statistics are equivalent to the selection-free test statistics if there is factually no selection into the intermediate transition conditional on $G$.
Formally, if the treated individuals are a randomly selected subset of all individuals at each level of $G$, we have the equality $\E(Y_d \mid G)=\E(Y \mid D=d, G)$, which implies the equivalence of the descriptive and selection-free test statistics.

The formulation of the descriptive tests is consistent with existing descriptive tests of the two theses in sociology.
When $\E(Y \mid D=d, G)$ is linear in $G$, $\Cov[\E(Y \mid D=d, G), G]/\Var(G)$ is the slope coefficient in the linear regression of $Y$ on $G$ in the subpopulation with $D=d$ \citep{torche_is_2011}.
When $\logit\{\E(D \mid G)\}$ and $\logit\{\E(Y \mid D=1, G)\}$ are linear in $G$, $\Cov \left\{ \logit[\E(D \mid G)], G \right\}/\Var(G)$ and $\Cov \left\{ \logit[\E(Y \mid D=1, G)], G \right\}/\Var(G)$ are respectively the slope coefficients in the logistic regression of $D$ on $G$ and in the logistic regression of $Y$ on $G$ in the subpopulation with $D=1$ \citep{mare_social_1980}.

\subsection{Literature review}
Previous studies have separately assessed the GE and ST theses in ways that are, to some extent, conceptually aligned with our selection-free tests. For the GE thesis, Zhou (\citeyear{zhou_equalization_2019}), Karlson (\citeyear{karlson_college_2019}), and Fiel (\citeyear{fiel_great_2020}) have all attempted to address the differential selection issue using various estimators, including residual balancing weighting, inverse probability weighting, and Bayesian regression models.
For the ST thesis, many authors have also designed various estimators in order to obtain selection-free tests \citep{mare_educational_1993, cameron_life_1998, lucas_neo-classical_2011, holm_dealing_2011, smith_late-stage_2019}.\footnote{\citet{mare_introduction_2011} also conceptually discusses two ways to construct selection-free tests (in his words, tests that correct for the ``unmeasured heterogeneity'') of the ST thesis which correspond to the deterministic and stochastic interpretations of our counterfactual slopes presented in Section~\ref{sec:proposal}. He failed to recognize that these two interpretations are, in fact, equivalent.}

However, these prior work all directly starts from a specific statistical estimator without formally defining the nonparametric causal estimand. This leads to three general problems. First, it is not always clear whether the selection-free tests of the prior work are compatible with one another and what assumptions are required for their estimators to provide valid selection-free tests.
When their findings differ even if the same data are used, one cannot immediately identify the reason why they differ, which could be any of three possibilities: different estimands, different identifying assumptions, and different estimation procedures.\footnote{\citet{zhou_equalization_2019} takes a more principled approach than other previous studies. He proposes a nonparametric estimand termed ``controlled mobility'', which conceptually corresponds to our counterfactual slopes. However, controlled mobility is not causally formulated in potential outcomes. Consequently, Zhou's (2019) test is conceptually less transparent.} Second, the lack of a nonparametric estimand prevents prior work from leveraging modern nonparametric theory in statistics to derive estimators that are flexible, robust, and efficient. We will return to this point in Section \ref{sec:estimation}.

Third, the lack of explicitly stated estimands and assumptions also prevents principled handling of confounders.
In fact, the statistical estimators in many previous studies are not well-interpretable due to the way they control for confounders.
Despite the variety and complexity of models used in the previous literature, a common problem with confounders can be intuitively illustrated by focusing on a simple additive linear regression:\footnote{This regression model is solely used to illustrate the problem of previous studies. We do not use such model in our own estimation.}
\begin{equation}
    \E(Y \mid D=1, G, \Z)=\alpha + \beta_1 G + \boldsymbol{\beta}_2^{\intercal} \Z, \label{equ:additive}
\end{equation}
where $\Z \defeq \X \setminus G$.
The regression-type estimators in the previous literature \citep{mare_educational_1993,smith_late-stage_2019, fiel_great_2020} essentially amounts to interpreting $\beta_1$ as a selection-free measure of the association between $Y$ and $G$ under $D=1$.
However, $\frac{\Cov(Y_1, G)}{\Var(G)}$, the natural building block for a selection-free test, cannot be captured by $\beta_1$ alone.\footnote{Under the additive linear model (\ref{equ:additive}) and the identifying assumptions in Section \ref{sec: assu}, $\E(Y_1 \mid G)=\alpha + \beta_1 G + \boldsymbol{\beta}_2^{\intercal} \E(\Z \mid G)$. Thus, $\frac{\Cov(Y_1, G)}{\Var(G)}=\beta_1 + \boldsymbol{\beta}_2^{\intercal} \frac{\Cov(Z, G)}{\Var(G)}$.}
Therefore, although previous work recognizes that a selection-free test can be constructed by leveraging confounders $\Z$, their estimates are not interpretable as selection-free tests.
In the context of the ST thesis, some researchers use bivariate probit models to account for selection into the lower-level transition \citep{lucas_neo-classical_2011, holm_dealing_2011}. These models rely on instrumental variables that are valid conditional on some covariates. Nevertheless, $\beta_1$ in equation (\ref{equ:additive}) is essentially still the focus of their interpretation, which again hinders the construction of a valid selection-free test.

There are also more idiosyncratic issues in the way some prior studies deal with covariates. In \citet{smith_late-stage_2019}, covariates measured after the intermediate transition (and even long after the outcome) are also controlled for, which renders the conditional ignorability of the transition untenable. In \citet{mare_educational_1993}, the identification of the latent class model relies on that, conditional on educational transitions, family background is independent of unobserved confounders. This is implausible, as family background and unobserved confounders both have effects on the transitions.
Finally,  \citet{mare_introduction_2011}, in a broad-stroke statement, contends that any variable affected by $G$ should not be controlled. However, this restriction is not only unnecessary but also detrimental to constructing a selection-free test, as controlling for post-$G$ covariates may be important for ensuring the conditional ignorability of $D$. 

The mishandling of covariates in the GE and ST literatures is reminiscent of mediation analysis before explicit estimands based on potential outcomes are introduced \citep{vanderweele_explanation_2015}. In our framework, we formally introduce the nonparametric causal estimands and their accompanying assumptions, which leads to an appropriate estimation procedure.
%Moreover, these works are based on parametric estimation, which is susceptible to model misspecification.

\section{Estimation}
\label{sec:estimation}
We develop estimators of the test statistics based on their efficient influence functions (EIF). The EIF is a central construct in modern non/semi-parametric statistics \citep{bickel_efficient_1998, van_der_vaart_asymptotic_2000, tsiatis_semiparametric_2006,kennedy_semiparametric_2022, hines_demystifying_2022}. In causal inference settings, an estimator constructed based on the EIF of the estimand has multiple advantages. First, the estimator is theoretically guaranteed to obtain optimal efficiency, i.e., its asymptotic variance is the lowest among all ``regular'' estimators. Second, it is typically multiply robust, which means that some constituent models can be misspecified while the final estimate remains consistent. Third, it typically allows for data-adaptive (e.g., Machine Learning [ML]) estimation while maintaining valid asymptotic inference \citep{van_der_laan_targeted_2011,chernozhukov_double/debiased_2018}. As a general approach that can be applied to a great variety of estimands, the EIF should be embraced by sociologists.\footnote{Apart from the current paper, \citet{zhou_attendance_2024} is another successful application of EIF-based estimation in answering sociological questions. }

In this section, we derive the EIFs of our slope estimands, construct the corresponding estimators, and discuss more specifically the desirable properties of our estimators. Previous tests of the GE and ST theses are conducted using estimators that do not have efficiency guarantee, are not robust to misspecification, and do not allow for principled data-adaptive estimation. Therefore, we also contribute to the GE and ST literatures by providing novel estimators with desirable properties.

\begin{prop}[EIF for $\xi_{\text{linear}}(d)$]
For the linear counterfactual slope $\xi_{\text{linear}}(d)$, the EIF is
     $$ \phi_{\text{linear}}(Y,d,\X) \defeq \frac{\left\{ \rho(Y,d,\X)-\E[\rho(Y,d,\X)] \right\} [G-\E(G)]}{\Var(G)} - \frac{G^2 - 2G\E(G) + [\E(G)]^2}{\Var(G)} \xi_{\text{linear}}(d), $$
     where $$\rho(Y,d,\X) \defeq \frac{\one(D=d)}{\Pro(D=d \mid \X)}[Y-\mu(d,\X)] + \mu(d,\X).$$
\end{prop}
The corresponding estimator is the solution of the estimating equation
$\frac{1}{n} \sum^n_{i=1} \phi_{\text{linear}}(Y_i,d,\X_i)=0$, i.e.,
\begin{equation*}
\hat{\xi}_{\text{linear}}(d) \defeq \frac{\frac{1}{n} \sum^n_{i=1} \hat{\rho}(Y_i,d,\X_i)G_i - \frac{1}{n} \sum^n_{i=1} \hat{\rho}(Y_i,d,\X_i) \frac{1}{n} \sum^n_{i=1} G_i}{\frac{1}{n} \sum^n_{i=1} G_i^2 - \left(\frac{1}{n} \sum^n_{i=1} G_i \right)^2},
\end{equation*}
where $\hat{\rho}(Y,d,\X)$ is the estimate of $\rho(Y,d,\X)$.
Note that the estimator $\hat{\xi}_{\text{linear}}(d)$ is the slope coefficient of the linear regression of $\hat{\rho}(Y,d,\X)$ on $G$. Intuitively, $\hat{\rho}(Y,d,\X)$ estimates $Y_d$ for each individual by pooling an imputation based on outcome regression with an imputation based on inverse probability weighting.

In $\hat{\xi}_{\text{linear}}(d)$, there are two nuisance functions, $\Pro(D=d \mid \X)$ and $\mu(d,\X)$. In our main analysis, we estimate these functions using parametric models. In particular, for $\Pro(D=d \mid \X)$, we use a logistic regression where $\X$ variables are entered additively; for $\mu(d,\X)$, we use a linear regression that contains all two-way interactions between $D$ and $\X$ as well as the constituent main effects. For both models, both a linear term and a squared term of $G$ are included. To stabilize the finite-sample performance of the estimators, we divide the initial estimate of $\one(D=d)/\Pro(D=d \mid \X)$ by its sample average.

The estimator $\hat{\xi}_{\text{linear}}(d)$ is doubly robust with respect to $\Pro(D=d \mid \X)$ and $\mu(d,\X)$. If either $\Pro(D=d \mid \X)$ or $\mu(d,\X)$ is consistently estimated, then $\xi_{\text{linear}}(d)$ is consistently estimated.\footnote{This double robustness property is due to the way $\rho(Y,d,\X)$ is constructed. For example, if $\hat{\Pro}(D=d \mid \X)$ converges to the true $\Pro(D=d \mid \X)$ but $\mu(d,\X)$ is misspecified such that it converges to an arbitrary $\Tilde{\mu}(d,\X)$, then $\E[\hat{\rho}(Y,d,\X)]$ converges to $\E \left\{\frac{\one(D=d)}{\Pro(D=d \mid \X)}[Y-\Tilde{\mu}(d,\X)] + \Tilde{\mu}(d,\X) \right\}=\E \left\{ \E \left\{\frac{\one(D=d)}{\Pro(D=d \mid \X)}[Y-\Tilde{\mu}(d,\X)] + \Tilde{\mu}(d,\X) \mid \X \right\} \right\} = \E[\mu(d,\X)-\Tilde{\mu}(d,\X)+\Tilde{\mu}(d,\X)]=\E[\mu(d,\X)].$ Conversely, if $\hat{\mu}(d,\X)$ converges to $\mu(d,\X)$ but $\hat{\Pro}(D=d \mid \X)$ does not converge to $\Pro(D=d \mid \X)$, $\E[\hat{\rho}(Y,d,\X)]$ also converges to $\E[\mu(d,\X)]$. The double robustness property of $\hat{\xi}_{\text{linear}}(d)$ and $\hat{\xi}_{\text{logit}}(d)$ can be proved by repeatedly applying similar arguments as well as the weak law of large numbers.}
When both $\Pro(D=d \mid \X)$ and $\mu(d,\X)$ are consistently estimated, $\hat{\xi}_{\text{linear}}(d)$ is asymptotically normal and semiparametrically efficient, i.e., $$\sqrt{n} \left( \hat{\xi}_{\text{linear}}(d) - \xi_{\text{linear}}(d) \right) \xrightarrow{} \mathcal{N} \left(0, \sigma^2_{\text{linear}}(d) \right),$$
where $\sigma^2_{\text{linear}}(d)=\E \left[\phi_{\text{linear}}(Y,d,\X)^2 \right]$ is the semiparametric efficiency bound for $\xi_{\text{linear}}(d)$. We can thus construct asymptotically accurate p-values and Wald-type confidence intervals for $\hat{\xi}_{\text{linear}}(d)$.

\begin{prop}[EIF for $\xi_{\text{logit}}(d)$]
For the logit counterfactual slope $\xi_{\text{logit}}(d)$, the EIF is
\begin{align*}
    &\phi_{\text{logit}}(Y,d,\X) \defeq \\
    &\frac{[G-\E(G)] \left\{ \frac{\rho(Y,d,\X)-\tau(d,G)}{\tau(d,G)[1-\tau(d,G)]} + \logit[\tau(d,G)] - \E\left\{ \logit[\tau(d,G)] \right\} \right\}}{\Var(G)} - \frac{G^2 - 2G\E(G) + [\E(G)]^2}{\Var(G)} \xi_{\text{logit}}(d),
\end{align*}
    where $\tau(d,G) \defeq \E(Y_d \mid G) = \E[\mu(d,\X) \mid G] = \E[\rho(Y, d,\X) \mid G]$.
\end{prop}
Similar to $\hat{\xi}_{\text{linear}}(d)$, the estimator corresponding to $\phi_{\text{logit}}(Y,d,\X)$ is
\vspace{-2em}
\begin{adjustwidth}{-0.2in}{0in}
    \begin{align*}
&\hat{\xi}_{\text{logit}}(d) \defeq \\
&\frac{\frac{1}{n} \sum^{n}_{i=1} \left\{\frac{\hat{\rho}(Y_i,d,\X_i)-\hat{\tau}(d,G_i)}{\hat{\tau}(d,G_i)[1-\hat{\tau}(d,G_i)]} +\logit[\hat{\tau}(d,G_i)] \right\} G_i - \frac{1}{n} \sum^{n}_{i=1} \left\{ \frac{\hat{\rho}(Y_i,d,\X_i)-\hat{\tau}(d,G_i)}{\hat{\tau}(d,G_i)[1-\hat{\tau}(d,G_i)]} +\logit[\hat{\tau}(d,G_i)] \right\} \times \frac{1}{n} \sum^{n}_{i=1} G_i}{\frac{1}{n} \sum^{n}_{i=1} G_i^2 - \left(\frac{1}{n} \sum^{n}_{i=1} G_i \right)^2},
\end{align*}
\end{adjustwidth}
where $\hat{\tau}(d,G)$ is the estimate of $\tau(d,G)$.
Note that $\hat{\xi}_{\text{logit}}(d)$ is the slope coefficient in the linear regression of $\frac{\hat{\rho}(Y,d,\X)-\hat{\tau}(d,G)}{\hat{\tau}(d,G)[1-\hat{\tau}(d,G)]} +\logit[\hat{\tau}(d,G)]$ on $G$.\footnote{In practice, the estimate of $\hat{\tau}(d,G)$ could fall outside of the $(0,1)$ range, which renders the logit transformation undefined. In that case, the researcher may censor $\hat{\tau}(d,G)$ at certain upper and lower bounds, for example, the largest and smallest estimated $\hat{\tau}(d,G)$ that are within the $(0,1)$ range. In our empirical test of the ST thesis, when $D$ is high school graduation, there are two individuals in the sample whose original $\hat{\tau}(1,G)$ is greater than 1. We thus recode $\hat{\tau}(1,G)$ to be 0.97 for these two individuals, which still makes them have the largest $\hat{\tau}(1,G)$ in the sample.}

In $\hat{\xi}_{\text{logit}}(d)$, in addition to $\Pro(D=d \mid \X)$ and $\mu(d,\X)$, there is a third nuisance function $\tau(d,G)$.
In our implementation, we use a linear regression for $\tau(d,G)$, where the dependent variable is the estimated function $\hat{\rho}(Y,d,\X)$, and the independent variables are  $G$ and its squared term.
The double robustness property of $\hat{\xi}_{\text{linear}}(d)$ also holds for $\hat{\xi}_{\text{logit}}(d)$, given that $\E[\hat{\rho}(Y,d,\X) \mid G]$ is consistently estimated.
When all three nuisance functions are consistently estimated, the estimator $\hat{\xi}_{\text{logit}}(d)$ is asymptotically normal and semiparametrically efficient, i.e., 
$$\sqrt{n} \left( \hat{\xi}_{\text{logit}}(d) - \xi_{\text{logit}}(d) \right) \xrightarrow{} \mathcal{N} \left(0, \sigma^2_{\text{logit}}(d) \right),$$
where $\sigma^2_{\text{logit}}(d)=\E \left[\phi_{\text{logit}}(Y,d,\X)^2 \right]$ is the semiparametric efficiency bound for $\xi_{\text{logit}}(d)$.
%The invoked condition is stronger than the condition for $\hat{\xi}_{\text{linear}}(d)$. As $\tau(d,G)$ is estimated by regressing $\hat{\rho}(Y,d,\X)$ on $G$, for $\sqrt{n}$-consistent estimation of $\tau(d,G)$ to hold, a sufficient set of conditions requires not only $\sqrt{n}$-consistent estimation of $\Pro(D=d \mid \X)$ and $\mu(d,\X)$ but also $\sqrt{n}$-consistent estimation of $\E[\hat{\rho}(Y,d,\X) \mid G]$.

In general, we make our estimation of the nuisance functions flexible with respect to interactions and nonlinearities. In doing so, we follow the philosophy of \citet{vansteelandt_assumption-lean_2022}, where the estimand is deliberately chosen to be parsimonious and interpretable, but estimation is allowed to be complex and flexible. In order to asses if our results are robust to data-adaptive estimation that can accommodate even more complicated functional forms, we also use ML to fit the nuisance functions. In Appendix \ref{appn:ml}, we present the details of the ML-based estimators and the empirical results. 

The proofs for Theorems 1 and 2 are presented in Appendix \ref{appn:proof}, where we also derive EIFs for the factual slopes. The estimators (and their asymptotic distributions) for the test statistics directly follow from those for the counterfactual and factual slopes.

\section{Empirical study}
\label{sec: empirical}
\subsection{Data and variables}
We apply our counterfactual slope approach to test the GE and ST theses using the National Longitudinal Survey of Youth 1979, which is a nationally representative dataset for a cohort of people who were born between 1957 and 1964 and were living in the United States at the baseline survey in 1979. We restrict the sample to respondents
who were between 14 to 17 years old in 1979 such that parental SES is measured before the intermediate transitions.

For both theses, parental SES ($G$) is measured as log family incomes averaged over the first three waves of the survey (1979, 1980, and 1981, when the respondents were 14 to 20 years old) and divided by the square root of family size to adjust for need \citep[e.g.,][]{zhou_equalization_2019}. For the GE thesis, the intermediate transition ($D$) is college graduation by age 31, and the outcome ($Y$) is log family incomes averaged over five survey waves between age 35 and 44, again divided by the square root of family size. For the ST thesis, we define three pairs of intermediate transition and outcome, corresponding to three separate analyses. First, high school graduation by 20 and college attendance by 23 are paired as $D$ and $Y$, the second pair is college attendance by 23 and college completion by 29, and the last pair is college completion by 29 and graduate degree completion by 34.

Baseline covariates used for causal identification ($\X$ variables) include gender, race, parental income and its squared term, mother's education, parental presence, the number of siblings, urban residence, educational expectation, friends’ educational expectation, significant other's educational expectation for the respondent, Armed Forces Qualification Test score, age at the baseline survey, the Rotter score of control locus, the Rosenberg self-esteem score, language spoken at home, separation from mother, Metropolitan Statistical Area category, whether a parent is foreign-born, region of residence, and mother’s working status.

For analyses where the intermediate transition is college completion, we drop individuals with parental income lower than 5000 dollars from the sample, as they have very low chances of completing college ($<$4.5\%), leading to overlap issues. For all analyses, we also delete individuals with missing values in any variables. Due to loss to follow up, most missing values are in income attainment, where 28\% are missing. For the educational transition indicators, missingness rates are 4\%, 11\%, 12\%, and 23\% for high school graduation, college attendance, college completion, and graduate degree attainment, respectively. Baseline covariates generally have few missing values, with the exception of parental presence (9\%).

\subsection{Findings}

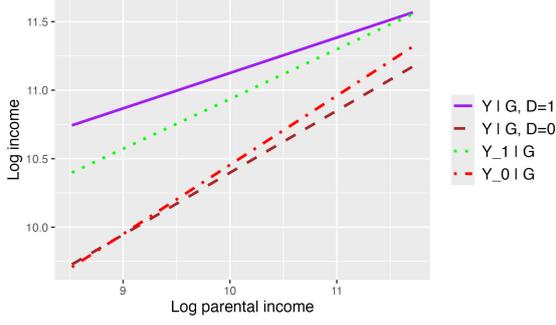
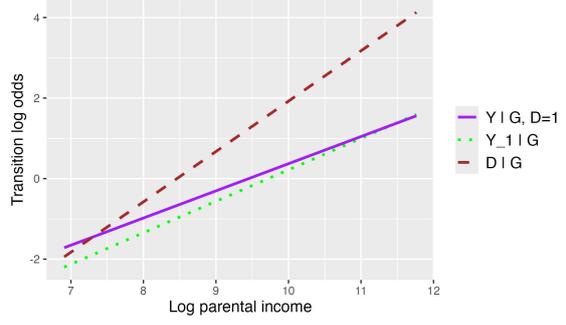
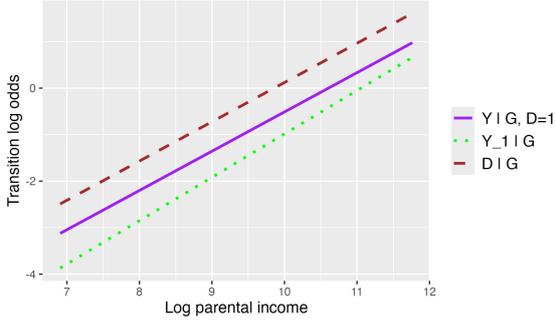
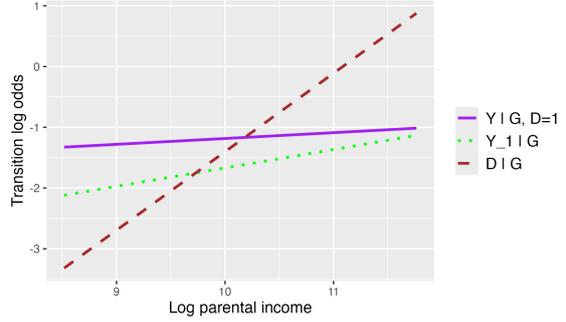
\begin{figure}[ht]
\centering

\subfigure[\small{GE: $D$=college degree, $Y$=income}]
{
\begin{tikzpicture}[scale = 1]
    \node at (0,0) {};
    \node at (1,1) {};

    \node[anchor=center, align=center, inner sep=0pt] at (0,0) {\includegraphics[width=.47\textwidth]{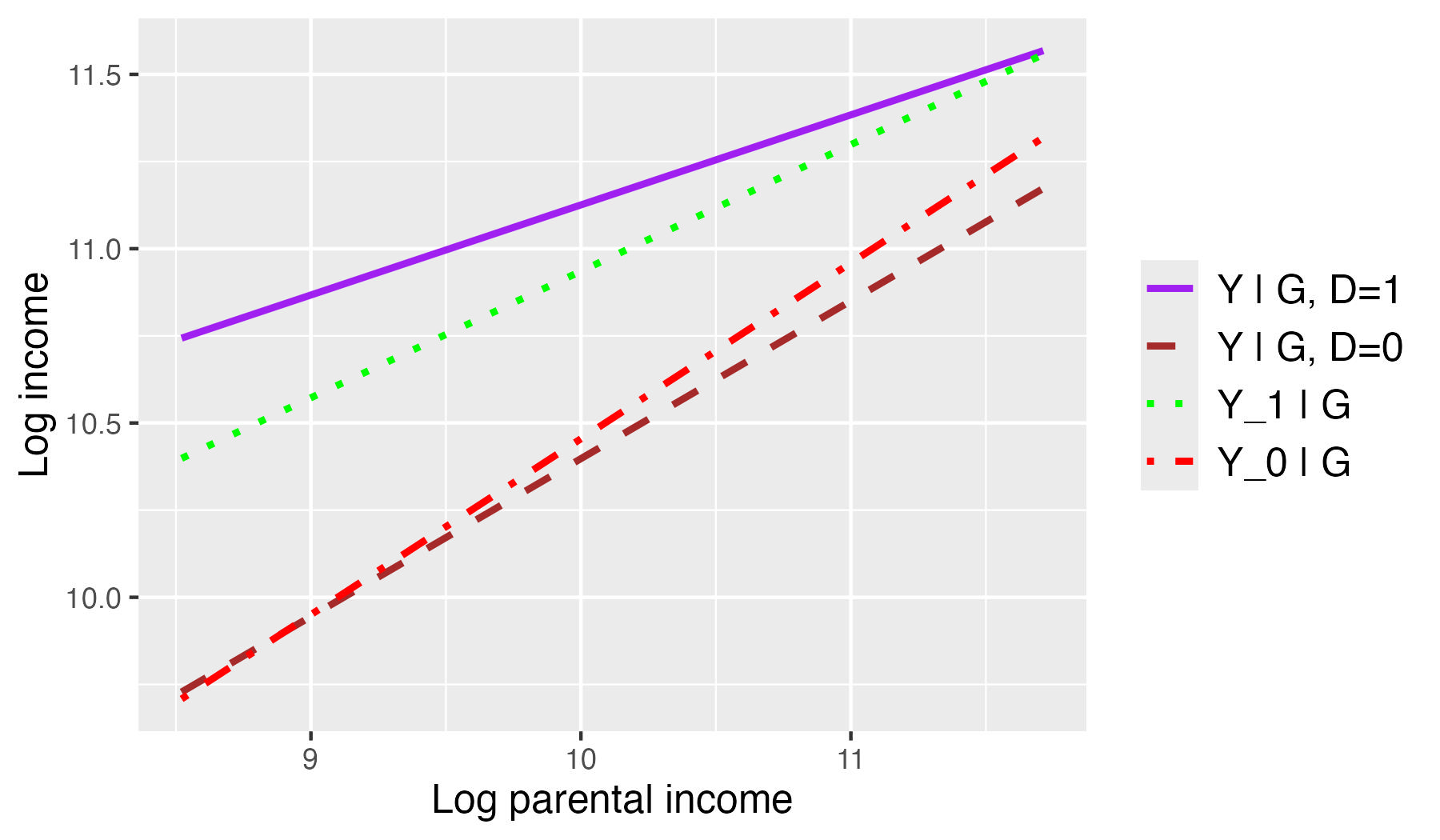}};
\end{tikzpicture}
}
\subfigure[\small{ST: $D$=high school, $Y$=college attendance}]
{
\begin{tikzpicture}[scale = 1]
    \node at (0,0) {};
    \node at (1,1) {};

    \node[anchor=center, align=center, inner sep=0pt] at (0,0) {\includegraphics[width=.47\textwidth]{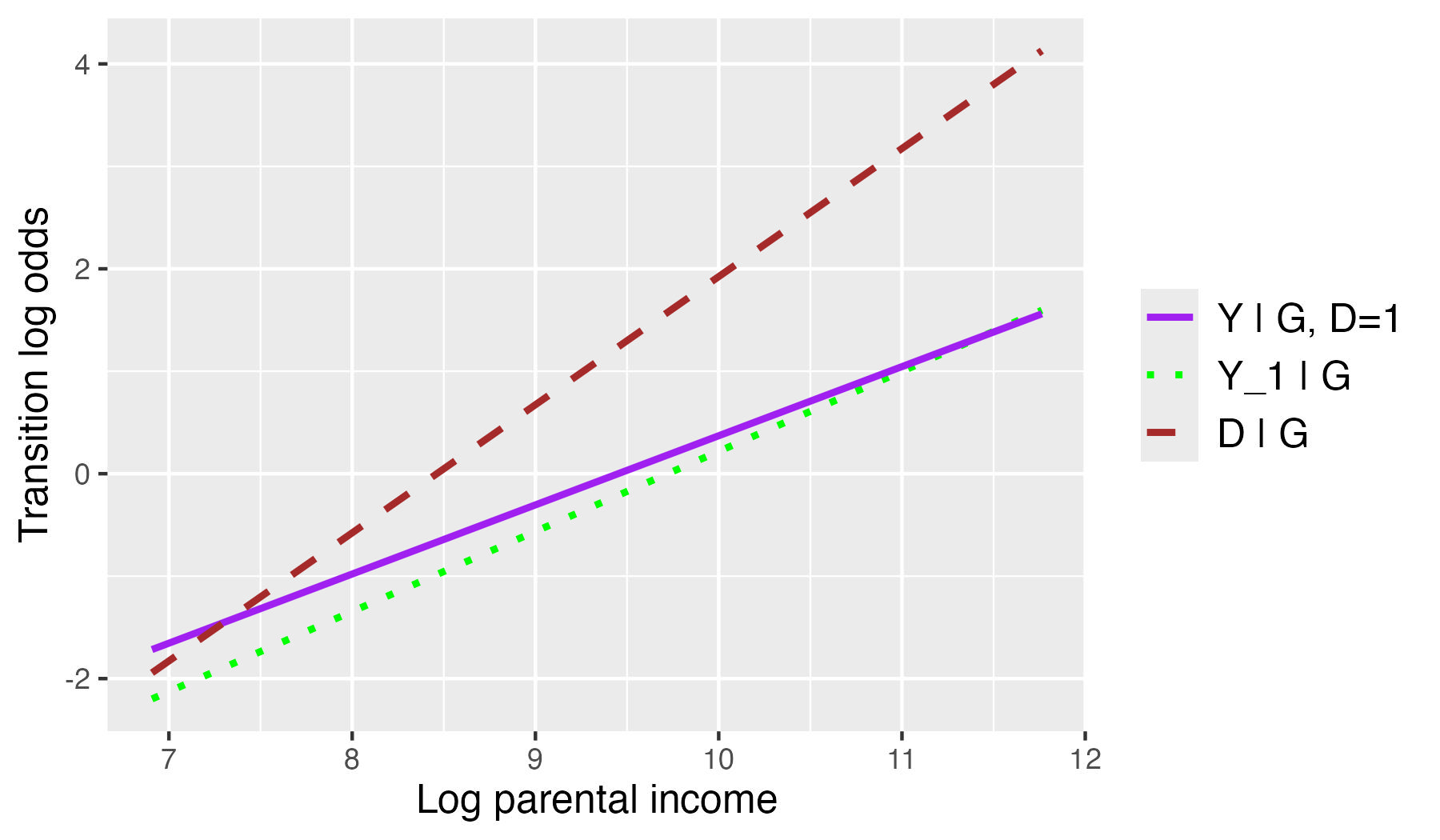}};
\end{tikzpicture}
}
\subfigure[\small{ST: $D$=college attendance, $Y$=college degree}]
{
\begin{tikzpicture}[scale = 1]
    \node at (0,0) {};
    \node at (1,1) {};

    \node[anchor=center, align=center, inner sep=0pt] at (0,0) {\includegraphics[width=.47\textwidth]{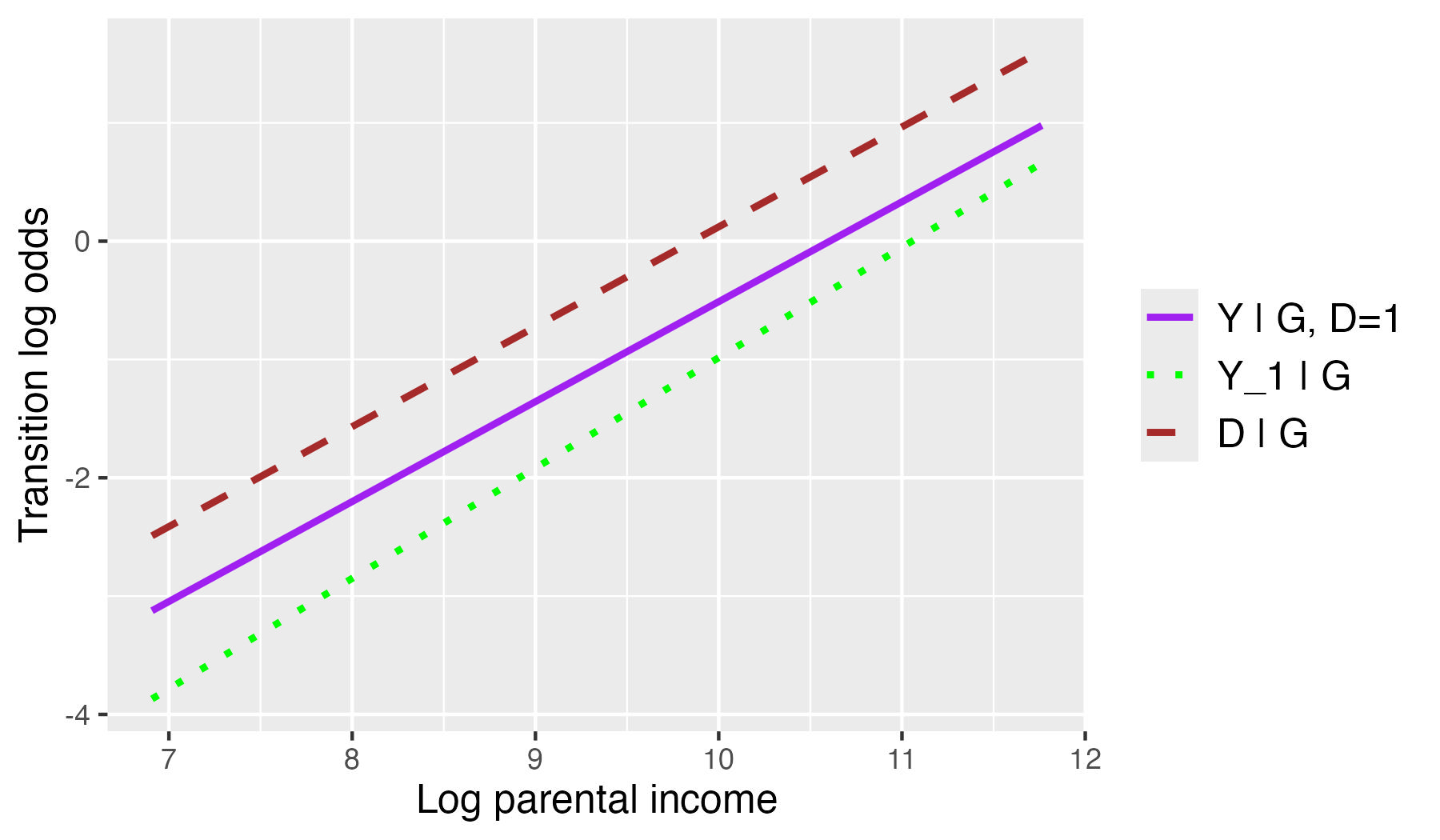}};
\end{tikzpicture}
}
\subfigure[\small{ST: $D$=college degree, $Y$=graduate degree}]
{
\begin{tikzpicture}[scale = 1]
    \node at (0,0) {};
    \node at (1,1) {};

    \node[anchor=center, align=center, inner sep=0pt] at (0,0) {\includegraphics[width=.47\textwidth]{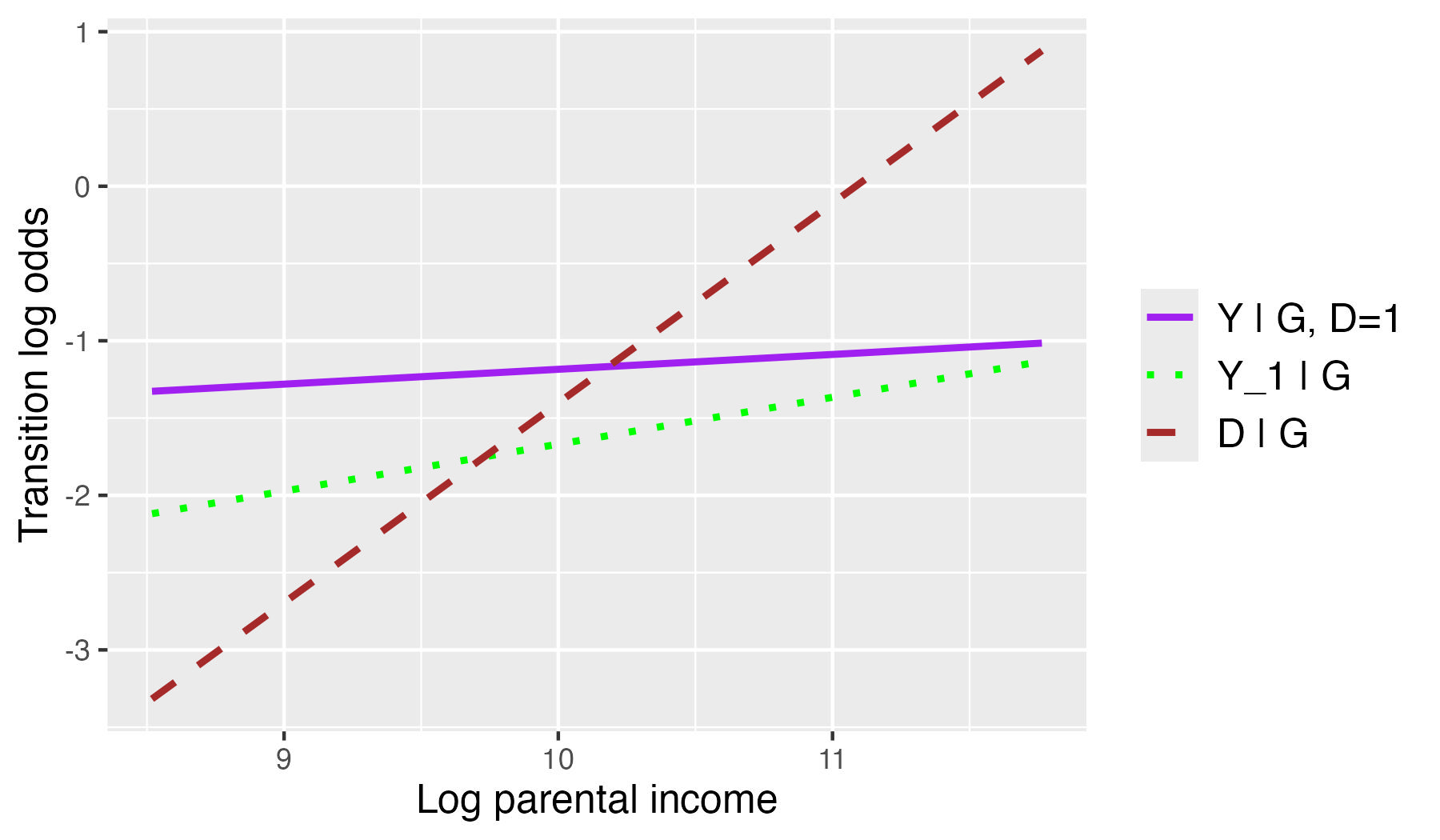}};
\end{tikzpicture}
}

\caption{Factual and counterfactual relationships between $G$ and $Y$. In Panels (a) and (d), the analyses are restricted to individuals with parental incomes higher than \$5000, who have a non-negligible rate of college completion. In Panel (a), for the lines representing $Y \mid G, D=d$, we plot the regression lines of $\hat{\rho}(Y,d,G)$ on $G$; for the lines representing $Y_d \mid G$, we plot the regression lines of $\hat{\rho}(Y,d,\X)$ on $G$. In Panel (b) through (d), we similarly plot regression lines using $\frac{\hat{\rho}(Y,1,\X)-\hat{\tau}(1,G)}{\hat{\tau}(1,G)[1-\hat{\tau}(1,G)]} +\logit [\hat{\tau}(1,G) ]$ for $Y_1 \mid G$, $\frac{\hat{\rho}(Y,1,G)-\hat{\E}(Y \mid D=1,G)}{\hat{\E}(Y \mid D=1,G)[1-\hat{\E}(Y \mid D=1,G)]} +\logit [\hat{\E}(Y \mid D=1,G) ]$ for $Y \mid G, D=1$, and $\frac{D-\hat{\E}(D \mid G)}{D-\hat{\E}(D \mid G) \left[D-\hat{\E}(D \mid G) \right]} + \logit [\hat{\E}(D \mid G) ]$ for $D \mid G$.} \label{fig:Result}
\end{figure}

In Figure \ref{fig:Result}, we plot the estimated factual and counterfactual slopes underlying the descriptive and selection-free tests.\footnote{It is also possible to plot the same relationships flexibly, using tools such as LOESS, if a more granular representation of the variable relationships is of interest. In the Github repository of our analysis, we provide code to do so.}
Tables \ref{tab:tests} and \ref{tab:estimates} respectively present the estimates of the test statistics and the underlying slope estimands. Descriptively, we confirm the GE thesis established in the literature, i.e., the association between parental income and income attainment conditional on having a college degree is weaker than the same association conditional on not having one. Hence, the estimate of the descriptive test statistic for the GE thesis is significantly positive. In Panel (a) of Figure \ref{fig:Result}, this is illustrated by the flatter slope of the line representing ``$Y \mid G, D=1$'' than the line representing ``$Y \mid G, D=0$''.
%However, in contrast to previous descriptive tests \citet{hout_status_1984, hout_more_1988, torche_is_2011}, we find that even among college graduates, SES origin is significantly associated with SES destination.

In selection-free terms, however, the GE thesis is not supported with sufficient evidence, as the selection-free test statistic is not statistically significant. In Panel (a) of Figure \ref{fig:Result}, this is shown by the fact that the line representing ``$Y_1 \mid G$'' and the line representing ``$Y_0 \mid G$'' have similar slopes. This is consistent with the selection-free tests conducted for the GE thesis by \citet{zhou_equalization_2019} and \citet{fiel_great_2020}. 

\begin{table}[h]
\caption{Estimates of descriptive and selection-free test statistics}
\centering
\begin{tabular}{llll}
  \hline
   & Descriptive test & Selection-free test  \\
  \hline
  GE: $D=$ college completion, $Y=$ income & $0.194^{**}$ & $0.142$ \\
  N=2711 & $(0.053,0.336)$ & $(-0.008, 0.291)$ \\
  ST: $D=$ high school, $Y=$ college attendance & $0.576^{***}$ & $0.469^{***}$ \\
  N=3603 & $(0.388, 0.763)$ & $(0.290, 0.648)$ \\
  ST: $D=$ college attendance, $Y=$ college degree & $0.000$ & $-0.088$  \\
  N=3402 & $(-0.196, 0.195)$ & $(-0.288, 0.111)$ \\
  ST: $D=$ college degree, $Y=$ graduate degree & $1.197^{***}$ & $0.989^{**}$ \\
  N=3082 & $(0.814, 1.579)$ & $(0.334, 1.644)$ \\
   \hline
\end{tabular}
\caption*{Note: P-value $<$ 0.01: **, $<$ 0.001: ***. 95 \% confidence intervals are in parentheses. The confidence intervals are Wald-type and constructed based on the asymptotically normal distributions of EIF-based estimators. Nuisance functions are estimated using parametric regression models.}
\label{tab:tests}
\end{table}

\begin{table}[h]
\caption{Estimates of factual and counterfactual slopes}
\centering
\begin{tabular}{llll}
  \hline
   & Slope & Confidence interval  \\
  \hline
  GE: $D=$ college completion, $Y=$ income &  &  \\
  $Y \mid G, D=1$ & $0.258^{***}$ & $(0.134, 0.383)$ \\
  $Y \mid G, D=0$ & $0.453^{***}$ & $(0.386, 0.520)$ \\
  $Y_1 \mid G$ & $0.363^{***}$ & $(0.226, 0.500)$ \\
  $Y_0 \mid G$ & $0.505^{***}$ & $(0.441, 0.568)$ \\
  ST: $D=$ high school, $Y=$ college attendance &  &  \\
  $D \mid G$ & $1.251^{***}$ & $(1.094, 1.407)$ \\
  $Y \mid G, D=1$ & $0.675^{***}$ & $(0.565, 0.785)$ \\
  $Y_1 \mid G$ & $0.781^{***}$ & $(0.663, 0.899)$ \\
  ST: $D=$ college attendance, $Y=$ college degree &  &  \\
  $D \mid G$ & $0.844^{***}$ & $(0.743, 0.945)$ \\
  $Y \mid G, D=1$ & $0.845^{***}$ & $(0.672, 1.017)$ \\
  $Y_1 \mid G$ & $0.933^{***}$ & $(0.742, 1.124)$ \\
  ST: $D=$ college degree, $Y=$ graduate degree &  &  \\
  $D \mid G$ & $1.293^{***}$ & $(1.129, 1.456)$ \\
  $Y \mid G, D=1$ & $0.096$ & $(-0.232, 0.424)$ \\
  $Y_1 \mid G$ & $0.304$ & $(-0.323, 0.930)$ \\
   \hline
\end{tabular}
\caption*{Note: P-value $<$ 0.01: **, $<$ 0.001: ***. The 95 \% confidence  intervals are Wald-type and constructed based on the asymptotically  normal distributions of EIF-based estimators. Nuisance functions are estimated using parametric regression models.}
\label{tab:estimates}
\end{table}

We can gain more insights about the differential selection process by comparing the four lines at different areas of the parental income distribution in Panel (a). In particular, only towards the lower end of the parental income distribution is the line for ``$Y \mid G, D=1$'' above the line for ``$Y_1 \mid G$''. This indicates that if selection into college completion were random, conditional on parental income, college graduates from more disadvantaged family backgrounds would obtain lower incomes in adulthood than they factually do. By implication, we can conclude that college graduates from lower income origins are selected (on characteristics positively associated with $Y_1$) in a way their counterparts from higher income origins are not. Conversely, only at the higher end of parental income distribution is the the line for ``$Y \mid G, D=0$'' below the line for ``$Y_0 \mid G$''. Hence, without differential selection into college completion, those who are from higher income origins but do not have a college degree would have higher incomes than they factually do. This implies that factually they are particularly selected on characteristics negatively associated with $Y_0$. On balance, the shift from the factual slopes to the counterfactual ones leads to a higher level of similarity between counterfactual slopes than between factual slopes.

%Consequently, both counterfactual slopes are steeper than their respective factual slopes. However, the particularly positive selection into college completion among individuals from lower income origin dominates the shift from the factual slopes to the counterfactual ones, leading to a higher level of similarity between counterfactual slopes than between factual slopes.

The descriptive ST thesis is confirmed by significantly positive test statistics for the first ($D=$ high school and $Y=$ college attendance) and the third ($D=$ college degree and $Y=$ graduate degree) pairs of transitions. However, the descriptive test statistic is estimated to be 0 for the second pair of transitions ($D=$ college attendance and $Y=$ college degree). 
%Hence, the descriptive association between family background and educational transitions declines from high school graduation to college attendance, remains constant from college attendance to college graduation, and declines again from college graduation to graduate degree attainment. 
%Conditional on having graduated college, the chance of obtaining a graduate degree in fact is no longer associated with parental income, as the factual slope is no longer significantly different from 0.

For the first and the third pairs of transitions, the ST thesis is also confirmed in selection-free tests, as the test statistics are still positive and significant. This lends support to the substantive interpretations of the ST thesis. For example, over educational transitions, perhaps the link between family background and aspiration for further education is attenuated \citep{stolzenberg_educational_1994}.
However, the magnitudes of these selection-free test statistics are smaller than the corresponding descriptive test statistics. Hence, the selection-free tests provide less support for the substantive interpretations of the ST thesis than the descriptive tests might otherwise appear to. This is similar to what we observe for the GE thesis and is indeed driven by a similar pattern of differential selection. 
In each of Panels (b) and (d) of Figure \ref{fig:Result}, comparing the lines representing ``$Y_1 \mid G$'' and ``$Y \mid G, D=1$'', the counterfactual slope is steeper than the factual slope. In particular, only at the lower end of parental income, students who have made the lower-level transition are positively selected on $Y_1$ and more likely to make the higher transition under the assignment of the lower-level transition than other students with a similar family background.
On the other hand, in Panel (c), although the counterfactual slope is also steeper than the factual slope, the contrast is not as marked.\footnote{Our tests of the ST thesis generally differ from previous ``selection-free'' tests in data, method, variables, as well as the measurement scale of the relationship between $Y$ and $G$. In terms of the changing relationship between $Y$ and $G$ over sequential transitions, our counterfactual slope estimates lead to conclusions different from those of \cite{cameron_life_1998} and \cite{smith_late-stage_2019}, who also let $G$ be family income.}

\FloatBarrier

\section{Alternative formulations of the ST tests}
We also consider two alternative formulations of the ST tests. First, the ST thesis can also be tested using the linear, instead of logit, slope estimands. Formally, the selection-free test statistic could be alternatively defined as 
\begin{equation}
    \frac{\Cov \left[ \E(D \mid G), G \right]}{\Var(G)} - \frac{\Cov \left[ \E(Y_1 \mid G), G \right]}{\Var(G)}. \label{equ:free_ST_linear}
\end{equation}
And the descriptive test statistic could be
\begin{equation}
    \frac{\Cov \left[ \E(D \mid G), G \right]}{\Var(G)} - \frac{\Cov \left[ \E(Y \mid D=1, G), G \right]}{\Var(G)}. \label{equ:desc_ST_linear}
\end{equation}

We estimate logit slopes in our main analysis because they follow the logit formulation of the original analysis of \citet{mare_social_1980} and illustrate the ability of our general approach to accommodate outcome transformations. On the other hand, the logit slope estimands take more complicated forms than the linear slope estimands, potentially reducing interpretability. Hence, we consider both pairs of tests reasonable formalizations of the ST thesis and present results on test statistics (\ref{equ:free_ST_linear}) and (\ref{equ:desc_ST_linear}) in Appendix \ref{appn:alternative}.

\begin{table}[h]
    \centering
    \caption{Variable definitions for three ST analyses}
    \begin{tabular}{l l l}
    \hline
        $P$ & $D$ & $Y$ \\
        \hline
        None & High school graduation & College attendance \\
        High school graduation & College attendance &  college degree \\
        College attendance &  college degree & Graduate degree completion \\
        \hline
    \end{tabular}
    \label{tab: def}
\end{table}

Second, in our main analysis, each ST test focuses on the intermediate transition ($D$) and the subsequent outcome transition ($Y$) without conditioning on the previous transition (denoted $P$ hereafter). The definitions of $P$, $D$, and $Y$ for each separate analysis of the ST thesis are summarized in Table \ref{tab: def}. Alternatively, the ST tests can be defined conditional on having made the transition $P$, i.e., conditional on $P=1$.  Formally, the selection-free test statistic could be 
\begin{equation}
    \frac{\Cov \left\{ \logit \left[\E(D \mid P=1, G) \right], G \right\}}{\Var(G)} - \frac{\Cov \left\{ \logit \left[\E(Y_1 \mid P=1, G) \right], G \right\}}{\Var(G)}, \label{equ:free_ST_cond}
\end{equation}
whose corresponding descriptive test statistic is\footnote{Note that $\E(Y \mid D=1, G) = \E(Y \mid P=1, D=1, G)$, because making the $P$ transition is a prerequisite for making the $D$ transition.} 
\begin{equation}
    \frac{\Cov \left\{ \logit[\E(D \mid P=1, G)], G \right\}}{\Var(G)} - \frac{\Cov \left\{ \logit[\E(Y \mid D=1, G)], G \right\}}{\Var(G)}. \label{equ:desc_ST_cond}
\end{equation}
We use test statistics (\ref{equ:free_ST}) and (\ref{equ:desc_ST}) in our main analysis for their structural affinity to the test statistics of the GE thesis. However, test statistic (\ref{equ:free_ST_cond}) has the advantage of making Assumption \ref{assu3} (overlap) more tenable, as its identification only requires $0 < \E(D \mid P=1, \x)<1$, not $0 < \E(D \mid \x) <1$ \citep[see][]{zhou_attendance_2024}. Given that both formulations are conceptually consistent with the narratives of the ST thesis, we consider them to be both reasonable. Again, we present the results based on the alternative formulation in Appendix \ref{appn:alternative}.

\section{Discussion}
In this paper, we re-evaluate two prominent theses in social stratification research by proposing and estimating a set of novel counterfactual slope estimands. The potential outcome formulation of these estimands allows us to conduct principled testing of the two theses in a selection-free manner. We develop robust and efficient estimators for our estimands based on efficient influence functions and apply our methods to a nationally representative dataset in the United States. Our selection-free tests show that the descriptive regularities of the two theses do not provide support for substantive interpretations of the theses as strongly as they might appear to.

Causal estimands have abounded in recent years and have found numerous applications in the social sciences \citep{imbens_causal_2024}. Our work showcases the power of the potential outcome framework in a non-standard setting. In the social world, self-selection is ubiquitous, which may complicate the interpretation of observed regularities. Using a carefully-constructed hypothetical intervention, one can formulate a counterfactual scenario where complication selection is absent. A selection-free test can then be constructed if the quantity of interest is identifiable in the counterfactual scenario. More broadly, this research demonstrates the potential of using thought experiments that are formalized as hypothetical interventions to adjudicate between competing theoretical explanations.

In this paper, we identify the counterfactual slope estimands using the conditional ignorability assumption. When unobserved confounders exist, it might also be possible to identify the estimands using alternative assumptions, which could be the focus of a future extension. This is an advantage of explicitly formulated causal estimands, which are not tied to either specific identifying assumptions nor specific estimators. Another possible extension in terms of identification is to incorporate time-varying covariates in the testing of the ST thesis over multiple educational transitions. For example, when the intermediate transition is college completion and the outcome is obtaining a graduate degree, covariates measured during college may aid the identification of the counterfactual slope. 

Apart from the two theses of theoretical interest in sociology, counterfactual slopes also provide a new framework for evaluating distributional impacts of policy interventions that assign a treatment $D$. The literature on distributional policy effects has focused on inequality measures defined solely in terms of the outcome itself, such as the variance or the Gini coefficient \citep[e.g.,][]{rothe_nonparametric_2010,firpo_identification_2016}. However, the inequality in an outcome ($Y$) indexed by a covariate ($G$) such as income is often also the focus of public concern and policy making. Thus, counterfactual slopes are useful for quantifying the $G$-based inequality in $Y$ under the assignment of treatment $D$ as mandated by a policy intervention. 

The counterfactual slopes in this paper represent the associations between the outcome and $G$ under hypothetical interventions constructed for testing the GE and ST theses. From a policy planning viewpoint, the definition of a counterfactual slope can be extended to accommodate  other interventions that are more relevant to policy practice. For example, one might be interested in estimating the counterfactual slope under an equalization intervention that leads to the same treatment rate across $G$ levels \citep[]{yu2023nonparametric} or an affirmative action-type intervention that targets the treatment rates at specific areas of the $G$ distribution.

\section*{Acknowledgements}
We are grateful to Guilherme Jardim Duarte, Felix Elwert, and three reviewers at Sociological Methodology for helpful comments. 
Earlier versions of this paper were presented at RC28 Spring Meeting in 2022 and the Annual Meeting of the American Sociological Association in 2023. We thank participants at these conferences for constructive discussions.

This research is partially supported by National Science Foundation through both NSF/NIGMS joint program (DMS-1953526/2122074) and statistics program (DMS-2310942), by National Institutes of Health (R01DC021431) as well as by the American Family Funding Initiative administered by the Data Science Institute of UW-Madison.

\section*{Appendices}
\begin{appendices}

\section{Derivation of EIFs for the main analysis}
\label{appn:proof} 
We use the Gateaux derivative approach to derive the EIFs \citep{ichimura_influence_2022}, which results in more succinct derivation than the approach traditionally used in the semiparametric causal inference literature \citep[e.g.,][]{hahn_role_1998}. To further simplify the derivation, we use some practical rules of calculating Gateaux derivatives \citep{hines_demystifying_2022, kennedy_semiparametric_2022}.

Let $\one_{\Tilde{o}}(o)$ be the point mass density at a single empirical observation,  $\Tilde{o}$. Let $\mathcal{P}_t$ denote a regular parametric submodel indexed by $t$. The subscript $t$ is omitted for the true model. By construction, $f_{\mathcal{P}_t}(o)=t \one_{\Tilde{o}}(o) + (1-t)f(o)$, i.e., the submodel is the true model perturbed in the direction of a single observation $\Tilde{o}$. Under this construction, the EIF of an estimand, $\xi$, is the Gateaux derivative at the truth, i.e., $\phi(\xi)=\frac{\partial \xi_{\mathcal{P}_t}}{\partial t} \big|_{t=0}$. For an arbitrary function $g(o)$, we denote $ \frac{\partial g_{\mathcal{P}_t}(o)}{\partial t} \big|_{t=0}$ as $\partial g(o)$.

\subsubsection*{Theorem 1, EIF of the linear counterfactual slope}

\begin{equation*}
    \xi_{\text{linear}}(d) = \frac{\E[G \mu(d,\X)]-\E[\mu(d,\X)]\E(G)}{\Var(G)} ,
\end{equation*}

We tackle this estimand part by part.
\begin{align*}
    &\phantom{{}={}} \partial \E_{\pt} [G \mu_{\pt}(d,\X)] \\
    &= \E[G \partial \mu_{\pt}(d,\X)] + \Tilde{g} \E(Y \mid d, \Tilde{\x}) - \E[G \mu(d,\X)] \\
    &= \E \left\{ G \frac{\one_{\Tilde{d}}(d) \one_{\Tilde{\x}}(\X)}{\Pro(D=d \mid \X)f(\X)} \left[\Tilde{y}-\mu(d,\X) \right] \right\} + \Tilde{g} \E(Y \mid d, \Tilde{\x}) - \E[G \mu(d,\X)] \\
    &= \int G \frac{\one_{\Tilde{d}}(d) \one_{\Tilde{\x}}(\x)}{\Pro(D=d \mid \x)} \left[\Tilde{y}-\mu(d,\x) \right] \dd \x + \Tilde{g} \E(Y \mid d, \Tilde{\x}) - \E[G \mu_d(\X)] \\
    &= \Tilde{g} \frac{\one_{\Tilde{d}}(d)}{\Pro(D=d \mid \Tilde{\x})} \left[\Tilde{y}-\mu(d,\Tilde{\x}) \right] + \Tilde{g} \E(Y \mid d, \Tilde{\x}) - \E[G \mu(d,\X)] \\
    &= G\rho(Y,d,\X) - \E[G \mu(d,\X)].
\end{align*}
\vspace{-3em}
\begin{align*}
     &\phantom{{}={}} \partial \{ \E[\mu(d,\X)]\E(G) \} \\
     &= \partial \E[\mu(d,\X)]\E(G) + \E[\mu(d,\X)] \partial \E(G) \\
     &= \{\rho(Y,d,\X)-\E[\mu(d,\X)]\}\E(G) + \E[\mu(d,\X)][G-\E(G)].
\end{align*}
\vspace{-3em}
\begin{align*}
 &\phantom{{}={}} \partial \Var(G) \\
 &= \partial \E(G^2) - 2\E(G) \partial \E(G)\\
 &= G^2 - 2G\E(G) + [\E(G)]^2 - \Var(G).
\end{align*}
Hence,
\vspace{-1em}
\begin{align*}
 &\phantom{{}={}} \phi_{\text{linear}}(d) \\
 &=  \frac{\left\{ \rho(Y,d,\X)-\E[\rho(Y,d,\X)] \right\} [G-\E(G)]}{\Var(G)} \\
 &\phantom{{}={}} - \frac{G^2 - 2G\E(G) + [\E(G)]^2 - \Var(G)}{[\Var(G)]^2} \{\E[G \mu(d,\X)]-\E[\mu(d,\X)]\E(G) \} - \xi_{\text{linear}}(d) \\
 &= \frac{\left\{ \rho(Y,d,\X)-\E[\rho(Y,d,\X)] \right\} [G-\E(G)]}{\Var(G)} - \frac{G^2 - 2G\E(G) + [\E(G)]^2}{\Var(G)} \xi_{\text{linear}}(d) .
\end{align*}

Finally, it is easy to see that the EIF for the factual slope $\frac{\Cov[ \E(Y \mid D=d, G), G ]}{\Var(G)}$ is 
$$\frac{\left\{ \rho(Y,d,\G)-\E[\rho(Y,d,\G)] \right\} [G-\E(G)]}{\Var(G)} - \frac{G^2 - 2G\E(G) + [\E(G)]^2}{\Var(G)} \frac{\Cov[ \E(Y \mid D=d, G), G ]}{\Var(G)},$$
\vspace{-1em}
where $$\rho(Y,d,G) \defeq \frac{\one(D=d)}{\Pro(D=d \mid G)}[Y-\E(Y \mid D=d, G)] + \E(Y \mid D=d, G).$$

\subsubsection*{Theorem 2, EIF of the logit counterfactual slope}

\begin{equation*}
    \xi_{\text{logit}}(d) = \frac{\E \left\{ G \log \frac{\E[\mu(d,\X) \mid G]}{1-\E[\mu(d,\X) \mid G]} \right\} - \E\left\{ \log \frac{\E[\mu(d,\X) \mid G]}{1-\E[\mu(d,\X) \mid G]} \right\}\E(G)}{\Var(G)}.
\end{equation*}
\vspace{-3em}
\begin{adjustwidth}{-1in}{0in}
\begin{align*}
    &\phantom{{}={}} \E \left\{G \partial \log \E_{\pt}[\mu_{\pt}(d,\X) \mid G] \right\} \\
    &= \int f(g) \frac{g}{\E[\mu(d,\X) \mid g]} \partial \int \mu_{\pt}(d,\x) \frac{f_{\pt}(\x,g)}{f_{\pt}(g)} \dd\x \dd g \\
    &= \int f(g) \frac{g}{\E[\mu(d,\X) \mid g]} \int \left\{ \partial \mu_{\pt}(d,\x) \frac{f(\x,g)}{f(g)} + \frac{\mu(d,\x)}{f(g)}[\one_{\Tilde{\x},\Tilde{g}}(\x,g)-f(\x,g)] - \frac{\mu(d,\x)f(\x,g)}{f(g)^2}[\one_{\Tilde{g}}(g)-f(g)] \right\} \dd\x \dd g \\
    &= \int \frac{g}{\E[\mu(d,\X) \mid g]} \int \left\{ \partial \mu_{\pt}(d,\x) f(\x,g) + \mu(d,\x)\one_{\Tilde{\x},\Tilde{g}}(\x,g) - \frac{\mu(d,\x)f(\x,g)}{f(g)}\one_{\Tilde{g}}(g) \right\} \dd\x \dd g \\
    &= \int \frac{g}{\E[\mu(d,\X) \mid g]} \int \left\{ \frac{\one_{\Tilde{d}}(d) \one_{\Tilde{\x}}(\x)}{\Pro(D=d \mid \x)} [\Tilde{y}-\mu(d,\x)] + \mu(d,\x)\one_{\Tilde{\x},\Tilde{g}}(\x,g) - \frac{\mu(d,\x)f(\x,g)}{f(g)}\one_{\Tilde{g}}(g) \right\} \dd\x \dd g \\
    &= \frac{G}{\E[\mu(d,\X) \mid G]} \left\{ \frac{\one(D=d)}{\Pro(D=d \mid \X)}[Y-\mu(d,\X)] + \mu(d,\X) \right\} - G.
\end{align*}
\end{adjustwidth}
Similarly,
\vspace{-1em}
\begin{align*}
    &\phantom{{}={}} \E \left\{G \partial \log \{1-\E_{\pt}[\mu_{\pt}(d,\X) \mid G] \}\right\} \\
    &= -\frac{G}{1-\E[\mu(d,\X) \mid G]} \left\{ \frac{\one(D=d)}{\Pro(D=d \mid \X)}[Y-\mu(d,\X)] + \mu(d,\X) \right\} + G \frac{\E[\mu(d,\X) \mid G]}{1-\E[\mu(d,\X) \mid G]}.
\end{align*}

Hence,
\begin{align*}
    &\phantom{{}={}} \partial \E_{\pt} \left\{ G \log \frac{\E_{\pt}[\mu_{\pt}(d,\X) \mid G]}{1-\E_{\pt}[\mu_{\pt}(d,\X) \mid G]} \right\} \\
    &= \E \left\{G \partial \log \frac{\E_{\pt}[\mu_{\pt}(d,\X) \mid G]}{1-\E_{\pt}[\mu_{\pt}(d,\X) \mid G]} \right\} + \Tilde{g} \log \frac{\E[\mu(d,\Tilde{\x}) \mid \Tilde{g}]}{1-\E[\mu(d,\Tilde{\x}) \mid \Tilde{g}]} - \E \left\{ G \log \frac{\E[\mu(d,\X) \mid G]}{1-\E[\mu(d,\X) \mid G]} \right\} \\
    &= \E \left\{G \partial \log \E_{\pt}[\mu_{\pt}(d,\X) \mid G] \right\} - \E \left\{G \partial \log \{1-\E_{\pt}[\mu_{\pt}(d,\X) \mid G] \}\right\} + \Tilde{g} \log \frac{\E[\mu(d,\Tilde{\x}) \mid \Tilde{g}]}{1-\E[\mu(d,\Tilde{\x}) \mid \Tilde{g}]} \\
    &\phantom{{}={}} - \E \left\{ G \log \frac{\E[\mu(d,\X) \mid G]}{1-\E[\mu(d,\X) \mid G]} \right\} \\
    &= G \frac{\rho(Y,d,\X)-\tau(d,G)}{\tau(d,G)[1-\tau(d,G)]} + G\log\frac{\tau(d,G)}{1-\tau(d,G)} - \E \left[ G\log\frac{\tau(d,G)}{1-\tau(d,G)} \right],
\end{align*}

Also,
\vspace{-2em}
\begin{adjustwidth}{-0.3in}{0in}
\begin{align*}
     &\phantom{{}={}} \partial \E_{\pt}\left\{ \log \frac{\E_{\pt}[\mu_{\pt}(d,X) \mid G]}{1-\E_{\pt}[\mu_{\pt}(d,X) \mid G]} \right\}\E_{\pt}(G) \\
     &= \E(G) \frac{\rho(Y,d,\X)-\tau(d,G)}{\tau(d,G)[1-\tau(d,G)]} + \E(G)\log\frac{\tau(d,G)}{1-\tau(d,G)} + G\E\left[ \log\frac{\tau(d,G)}{1-\tau(d,G)} \right] - 2 \E(G)\E\left[ \log\frac{\tau(d,G)}{1-\tau(d,G)} \right].
\end{align*}
\end{adjustwidth}
Finally, we obtain
\vspace{-2em}
\begin{adjustwidth}{-0.4in}{0in}
\begin{align*}
    &\phantom{{}={}} \phi_{\text{logit}}(d) \\
    &= \frac{[G-\E(G)] \left\{ \frac{\rho(Y,d,\X)-\tau(d,G)}{\tau(d,G)[1-\tau(d,G)]} +\log\frac{\tau(d,G)}{1-\tau(d,G)} - \E\left[ \log\frac{\tau(d,G)}{1-\tau(d,G)} \right] \right\}}{\Var(G)} - \xi_{\text{logit}}(d) - \\
    &\phantom{{}={}}  \frac{G^2 - 2G\E(G) + [\E(G)]^2 - \Var(G)}{[\Var(G)]^2} \left\{ \E \left\{ G \log \frac{\E[\mu(d,\X) \mid G]}{1-\E[\mu(d,\X) \mid G]} \right\} - \E\left\{ \log \frac{\E[\mu(d,\X) \mid G]}{1-\E[\mu(d,\X) \mid G]} \right\}\E(G) \right\} \\
    &= \frac{[G-\E(G)] \left\{ \frac{\rho(Y,d,\X)-\tau(d,G)}{\tau(d,G)[1-\tau(d,G)]} +\log\frac{\tau(d,G)}{1-\tau(d,G)} - \E\left[ \log\frac{\tau(d,G)}{1-\tau(d,G)} \right] \right\}}{\Var(G)} - \frac{G^2 - 2G\E(G) + [\E(G)]^2}{\Var(G)} \xi_{\text{logit}}(d).
\end{align*}
\end{adjustwidth}

Now we turn to the factual slopes. The EIF for
$\frac{\Cov \left[ \log\frac{\E(D \mid G)}{1-\E(D \mid  G)}, G \right]}{\Var(G)}$ is
\begin{align*}
    & \frac{[G-\E(G)] \left\{ \frac{D-\E(D \mid G)}{\E(D \mid G)[1-\E(D \mid G)]} +\log\frac{\E(D \mid G)}{1-\E(D \mid G)} - \E\left[ \log\frac{\E(D \mid G)}{1-\E(D \mid G)} \right] \right\}}{\Var(G)} \\
    &- \frac{G^2 - 2G\E(G) + [\E(G)]^2}{\Var(G)} \frac{\Cov \left[ \log\frac{\E(D \mid G)}{1-\E(D \mid  G)}, G \right]}{\Var(G)}.
\end{align*}

And the EIF for $\frac{\Cov \left[ \log\frac{\E(Y \mid D=1, G)}{1-\E(Y \mid D=1, G)}, G \right]}{\Var(G)}$ is
\begin{align*}
    & \frac{[G-\E(G)] \left\{ \frac{\rho(Y,1,G)-\E(Y \mid D=1,G)}{\E(Y \mid D=1,G)[1-\E(Y \mid D=1,G)]} +\log\frac{\E(Y \mid D=1,G)}{1-\E(Y \mid D=1,G)} - \E\left[ \log\frac{\E(Y \mid D=1,G)}{1-\E(Y \mid D=1,G)} \right] \right\}}{\Var(G)} \\
    &- \frac{G^2 - 2G\E(G) + [\E(G)]^2}{\Var(G)} \frac{\Cov \left[ \log\frac{\E(Y \mid D=1, G)}{1-\E(Y \mid D=1, G)}, G \right]}{\Var(G)}.
\end{align*}

\begin{comment}
\section{Proof of double robustness}
We first prove the double robustness of $\hat{\xi}_{\text{linear}}(d)$.

\begin{equation*}
\hat{\xi}_{\text{linear}}(d) \defeq \frac{\frac{1}{n} \sum^n_{i=1} \left\{\frac{\one(D_i=d)}{\hat{\Pro}(D=d \mid \X_i)}[Y-\hat{\mu}(d,\X_i)] + \hat{\mu}(d,\X_i) \right\} G_i - \frac{1}{n} \sum^n_{i=1} \left\{\frac{\one(D_i=d)}{\hat{\Pro}(D=d \mid \X_i)}[Y-\hat{\mu}(d,\X_i)] + \hat{\mu}(d,\X_i) \right\} \frac{1}{n} \sum^n_{i=1} G_i}{\frac{1}{n} \sum^n_{i=1} G_i^2 - \left(\frac{1}{n} \sum^n_{i=1} G_i \right)^2},
\end{equation*}
\end{comment}

\section{Machine learning implementation of the estimators}
\label{appn:ml}
In the main body of the paper, we fit nuisance functions in the estimators for the test statistics using parametric regression models. As a robustness check, we present a set of results here, where the nuisance functions are fit using nonparametric ML models. In particular, we use Neural Networks with a single hidden layer whose size and decay parameters are chosen via cross-validation. The use of ML allows our estimation to be free of functional form assumptions and hence free of specification errors that might inflict parametric models.

To improve the validity of asymptotic inference when ML is used to fit the nuisance functions, we apply cross-fitting \citep{chernozhukov_double/debiased_2018, diaz_machine_2020, kennedy_semiparametric_2022}: we randomly split the sample into two subsamples, fit the nuisance functions in subsample one, then plug in variable values from sample two to get fitted values, and finally this procedure is repeated with the roles of the two subsamples swapped. Our procedure is an instance of the Double/debiased ML framework \citep{chernozhukov_double/debiased_2018, chernozhukov_doubledebiasedneyman_2017}.

Recall that in $\hat{\xi}_{\text{linear}}(d)$, there are two nuisance functions, $\Pro(D=d \mid \X)$ and $\mu(d,\X)$.
And in $\hat{\xi}_{\text{logit}}(d)$, there is additionally a third nuisance function, $$\tau(d,G) \defeq \E[\rho(Y,d,\X) \mid G] \defeq \E \left\{\frac{\one(D=d)}{\Pro(D=d \mid \X)}[Y-\mu(d,\X)] + \mu(d,\X) \mid G \right\}.$$ The cross-fitting procedure for $\tau(d,G)$ is somewhat more involved: 
We again first randomly split the sample into two subsamples. Then we estimate $\Pro(D=d \mid \X)$ and $\mu(d,\X)$ in each subsample \emph{without} cross-fitting and obtain $\hat{\rho}(Y,d,\X)$. Finally, we obtain $\hat{\tau}(d,G)$ using cross-fitting, i.e., we fit $\E\left[\hat{\rho}(Y,d,\X) \mid G \right]$ separately in each subsample and plug in values of $G$ from the respective other subsample. Using this procedure, we ensure that the fitting of $\tau(d,G)$ is done separately in each subsample.\footnote{In terms of estimation and inference, there is a similar nuisance function in the conditional decomposition of \citet{yu2023nonparametric}. We refer readers to that paper for additional technical discussion.} 

\begin{table}[ht]
\caption*{Table B1. Estimates of descriptive and selection-free test statistics based on Neural Networks}
\centering
\begin{tabular}{llll}
  \hline
   & Descriptive test & Selection-free test  \\
  \hline
  GE: $D=$ college degree, $Y=$ income & $0.194^{**}$ & $0.111$ \\
  N=2711 & $(0.054,0.335)$ & $(-0.019, 0.240)$ \\
  ST: $D=$ high school, $Y=$ college attendance & $0.535^{***}$ & $0.389^{***}$ \\
  N=3603 & $(0.377, 0.692)$ & $(0.240, 0.538)$ \\
  ST: $D=$ college attendance, $Y=$ college degree & $0.413$ & $0.172$  \\
  N=3402 & $(-0.272, 1.099)$ & $(-0.338, 0.683)$ \\
  ST: $D=$ college degree, $Y=$ graduate degree & $1.121^{***}$ & $1.163^{***}$ \\
  N=3082 & $(0.745, 1.497)$ & $(0.486, 1.839)$ \\
   \hline
\end{tabular}
   \caption*{Note: P-value $<$ 0.01: **, $<$ 0.001: ***. Confidence intervals are in parentheses. \\ Cross-fitting is applied. The confidence intervals are Wald-type and constructed \\ based on the asymptotically normal distributions of EIF-based estimators.}
\end{table}

The empirical results based on ML models are presented in Table B1. In terms of statistical significance, the results are entirely consistent with the results based on parametric regression models. For all significant estimates, the signs of the estimates also do not change.

\FloatBarrier

\begin{comment}
    \begin{table}[h]
\centering
\caption{Estimates of factual and counterfactual slopes based on Neural Networks}
\begin{tabular}{llll}
  \hline
   & Slope & Confidence interval  \\
  \hline
  GE: $D=$ college degree, $Y=$ income &  &  \\
  $Y \mid G, D=1$ & $0.260^{***}$ & $(0.136, 0.384)$ \\
  $Y \mid G, D=0$ & $0.454^{***}$ & $(0.388, 0.520)$ \\
  $Y_1 \mid G$ & $0.396^{***}$ & $(0.281, 0.510)$ \\
  $Y_0 \mid G$ & $0.507^{***}$ & $(0.443, 0.571)$ \\
  ST: $D=$ high school, $Y=$ college attendance &  &  \\
  $D \mid G$ & $1.181^{***}$ & $(1.063, 1.299)$ \\
  $Y \mid G, D=1$ & $0.646^{***}$ & $(0.531, 0.761)$ \\
  $Y_1 \mid G$ & $0.792^{***}$ & $(0.675, 0.908)$ \\
  ST: $D=$ college attendance, $Y=$ college degree &  &  \\
  $D \mid G$ & $0.837^{***}$ & $(0.734, 0.940)$ \\
  $Y \mid G, D=1$ & $0.423$ & $(-0.281, 1.127)$ \\
  $Y_1 \mid G$ & $0.664^{*}$ & $(0.143, 1.186)$ \\
  ST: $D=$ college degree, $Y=$ graduate degree &  &  \\
  $D \mid G$ & $1.274^{***}$ & $(1.121, 1.427)$ \\
  $Y \mid G, D=1$ & $0.153$ & $(-0.186, 0.492)$ \\
  $Y_1 \mid G$ & $0.111$ & $(-0.562, 0.785)$ \\
   \hline
   \multicolumn{3}{l}{\makecell{Note: P-value $<$ 0.01: **, $<$ 0.001: ***. Cross-fitting is applied. The confidence \\ intervals are Wald-type and constructed based on the asymptotically normal \\ distributions of EIF-based estimators.}}
\end{tabular}
\label{tab:estimates_ml}
\end{table}
\end{comment}

\section{Alternative tests of the ST thesis}
\label{appn:alternative} 

Compared with the main analysis, we need to estimate two new estimands for the alternative ST tests: $\frac{\Cov \left( D, G \right)}{\Var(G)}$ and 
$\frac{\Cov \left\{ \logit \left[\E(Y_1 \mid P=1, G) \right], G \right\}}{\Var(G)}$. 

\begin{assu}[Conditional ignorability, alternative form]
     $Y_d \indep D \mid P=1, \X=\x, \forall d, \x$. \label{assu4}
\end{assu}
\vspace{-2em}
\begin{assu}[Overlap, alternative form]
    $0 < \E(D \mid P=1, \x) <1, \forall \x$.  \label{assu5}
\end{assu}

First, under Assumptions \ref{assu2}, \ref{assu4}, and  \ref{assu5}, we have the following identification result
\begin{align*}
   &\phantom{{}={}} \frac{\Cov \left\{ \logit \left[\E(Y_1 \mid P=1, G) \right], G \right\}}{\Var(G)} \\
   &=\frac{\E \left\{ G \cdot  \logit \left[\E(\mu(1,X) \mid P=1, G) \right] \right\} - \E\{\logit \left[\E(\mu(1,X) \mid P=1, G) \right] \} \E(G) }{\Var(G)}.
\end{align*}

Next, we derive the EIFs. The detailed steps are similar to what appear in Appendix \ref{appn:proof} and hence are omitted. 

The EIF of $\frac{\Cov \left( D, G \right)}{\Var(G)}$ is
$$\frac{[ D-\E(D) ] [G-\E(G)]}{\Var(G)} - \frac{G^2 - 2G\E(G) + [\E(G)]^2}{\Var(G)} \frac{\Cov \left( D, G \right)}{\Var(G)} .$$

And the EIF for $\ddot{\xi} \defeq \frac{\Cov \left\{ \logit \left[\E(Y_1 \mid P=1, G) \right], G \right\}}{\Var(G)}$ is
$$\frac{[G-\E(G)] \left\{ \frac{\rho(Y,1,1,\X)-\tau(1,1,G)}{\tau(1,1,G)[1-\tau(1,1,G)]} +\log\frac{\tau(1,1,G)}{1-\tau(1,1,G)} - \E\left[ \log\frac{\tau(1,1,G)}{1-\tau(1,1,G)} \right] \right\}}{\Var(G)} - \frac{G^2 - 2G\E(G) + [\E(G)]^2}{\Var(G)} \ddot{\xi},$$
where $$\rho(Y,1,1,\X) \defeq \frac{\one(D=1)}{\Pro(D=1 \mid P=1, \X)}[Y-\mu(1,\X)] + \mu(1,\X),$$
and $\tau(1,1,G) \defeq \E(Y_1 \mid P=1, G) = \E[\mu(1,\X) \mid P=1,G] = \E[\rho(Y, 1,1,\X) \mid P=1, G]$.

\begin{table}[h]
\centering
\caption*{Table C1. Estimates of test statistics for the ST thesis based on linear slope estimands}
\begin{tabular}{llll}
  \hline
   & Descriptive test & Selection-free test  \\
  \hline
  $D=$ high school, $Y=$ college attendance & $0.005$ & $-0.018$ \\
  N=3603 & $(-0.025, 0.035)$ & $(-0.046, 0.011)$ \\
  $D=$ college attendance, $Y=$ college degree & $0.009$ & $0.022$  \\
  N=3402 & $(-0.027, 0.044)$ & $(-0.007, 0.052)$ \\
  $D=$ college degree, $Y=$ graduate degree & $0.179^{***}$ & $0.163^{***}$ \\
  N=3082 & $(0.115, 0.244)$ & $(0.090, 0.236)$ \\
   \hline
\end{tabular}
   \caption*{Note: P-value $<$ 0.01: **, $<$ 0.001: ***. 95 \% confidence intervals are in parentheses. The confidence intervals are Wald-type and constructed based on the asymptotically normal distributions of EIF-based estimators. Nuisance functions are estimated using parametric regression models.}
\end{table}

\begin{table}[h]
\caption*{Table C2. Estimates of test statistics for the ST thesis based on slope estimands conditional on making the previous transition ($P=1$)}
\centering
\begin{tabular}{llll}
  \hline
   & Descriptive test & Selection-free test  \\
  \hline
  $D=$ college attendance, $Y=$ college degree & $-0.168$ & $-0.226$  \\
  N=3396 & $(-0.368, 0.032)$ & $(-0.402, -0.050)$ \\
  $D=$ college degree, $Y=$ graduate degree & $0.942^{***}$ & $0.869^{***}$ \\
  N=2969 & $(0.554, 1.330)$ & $(0.468, 1.270)$ \\
   \hline
\end{tabular}
   \caption*{Note: P-value $<$ 0.01: **, $<$ 0.001: ***. 95 \% confidence intervals are in parentheses. The confidence intervals are Wald-type and constructed based on the asymptotically normal distributions of EIF-based estimators. Nuisance functions are estimated using parametric regression models. The results are omitted for the case where $D$ is high school graduation, and $Y$ is college attendance, because there is no $P$ in that case.}
\end{table}

In Table C1, we present estimates of test statistics (\ref{equ:free_ST_linear}) and (\ref{equ:desc_ST_linear}), which are based on linear slope estimands. 
When the intermediate transition ($D$) is college completion, and the outcome ($Y$) is the attainment of a graduate degree, the conclusion remains qualitatively the same as that in the main analysis. When $D$ is college attendance, and $Y$ is college completion, both test statistics remain statistically insignificant, which is again consistent with the main analysis. However, when $D$ is high school graduation, and $Y$ is college attendance, we do not have descriptive evidence in support of the ST thesis anymore, which contradicts the main analysis.

In Table C2, we present estimates of test statistics (\ref{equ:free_ST_cond}) and (\ref{equ:desc_ST_cond}), which are based on slope estimands conditional on the previous transition $P$. 
When the intermediate transition ($D$) is college completion, and the outcome ($Y$) is the attainment of a graduate degree, the  conclusion again remains qualitatively the same as that in the main analysis. When $D$ is college attendance, and $Y$ is college completion, both test statistics again remain statistically insignificant. Finally, the results are omitted for the case where $D$ is high school graduation, and $Y$ is college attendance, because there is no $P$ in that case, which means that the estimates are the same as those in the main analysis by definition.

In Tables C1 and C2, the selection-free test statistics are smaller than the descriptive test statistics in four out of five pairs of comparison. This attests to the robustness of the differential selection pattern identified in the main analysis to alternative definitions of the slope estimands.

\end{appendices}

\FloatBarrier

\bibliographystyle{asr}
\bibliography{references}

\end{document}